\documentclass[twocolumn,twocolappendix]{aastex63}

\usepackage{amsmath}
\usepackage{bm} 
\usepackage{graphicx}
\usepackage[caption=false]{subfig} 
\usepackage{ulem} 

\usepackage{acronym}
\newacro{DF}{distribution function}
\newacroplural{DFs}{distribution functions}
\newcommand{\DF}{\ac{DF}}
\newcommand{\DFs}{\acp{DF}}
\newacro{MW}{Milky Way}
\newcommand{\MW}{\ac{MW}}
\newacro{LMC}{Large Magellanic Cloud}
\newcommand{\LMC}{\ac{LMC}}
\newacro{SMC}{Small Magellanic Cloud}
\newcommand{\SMC}{\ac{SMC}}
\newacro{CBE}{collisionless Boltzmann equation}
\newcommand{\CBE}{\ac{CBE}}
\newacro{DM}{Dark Matter}
\newcommand{\DM}{\ac{DM}}
\newacro{COM}{centre of mass}
\newcommand{\COM}{\ac{COM}}

\newcommand{\ri}{\mathrm{i}}

\newcommand{\re}{\mathrm{e}}
\newcommand{\rd}{\mathrm{d}}

\newcommand{\rs}{\mathrm{s}}

\newcommand{\rc}{\mathrm{c}}
\newcommand{\bT}{\bm{\theta}}
\newcommand{\bJ}{\mathbf{J}}
\newcommand{\bx}{\mathbf{x}}
\renewcommand{\bv}{\mathbf{v}}

\newcommand{\bn}{\mathbf{n}}

\newcommand{\tbJ}{\widetilde{\mathbf{J}}}
\newcommand{\tbn}{\widetilde{\mathbf{n}}}
\newcommand{\bO}{\mathbf{\Omega}}
\newcommand{\tbO}{\widetilde{\mathbf{\Omega}}}

\newcommand{\bM}{\mathbf{M}}
\newcommand{\ubM}{\underline{\mathbf{M}}}
\newcommand{\bN}{\mathbf{N}}
\newcommand{\ubI}{\underline{\mathbf{I}}}

\newcommand{\rp}{r_{\mathrm{p}}}
\newcommand{\ra}{r_{\mathrm{a}}}

\newcommand{\bA}{\mathbf{A}}

\newcommand{\mC}{\mathcal{C}}

\newcommand{\Op}{\Omega_{\mathrm p}}
\newcommand{\op}{\omega_{\mathrm p}}

\newcommand{\ba}{\mathbf a}
\newcommand{\bb}{\mathbf b}
\newcommand{\bc}{\mathbf c}

\newcommand{\uba}{\underline{\mathbf a}}
\newcommand{\ubb}{\underline{\mathbf b}}
\newcommand{\ubas}{\underline{\mathbf a}_{\mathrm s}}
\newcommand{\ubab}{\underline{\mathbf a}_{\mathrm b}}

\newcommand{\OILR}{\Omega_{\mathrm{ILR}}}

\newcommand{\Rb}{R_{\mathrm b}}

\newcommand{\nmax}{n_{\mathrm{max}}}

\newcommand{\Ra}{R_{\mathrm a}}

\newcommand{\gtorder}{\mathrel{\raise.3ex\hbox{$>$}\mkern-14mu
             \lower0.6ex\hbox{$\sim$}}}
\newcommand{\ltorder}{\mathrel{\raise.3ex\hbox{$<$}\mkern-14mu
             \lower0.6ex\hbox{$\sim$}}}

\newcommand{\Mtot}{M_{\mathrm{tot}}}

\newcommand{\MMW}{M_{\mathrm{MW}}}
\newcommand{\MDM}{M_{\mathrm{DM}}}
\newcommand{\FDM}{F_{\mathrm{DM}}}

\newcommand{\ubaDM}{\uba_{\mathrm{DM}}}

\newcommand{\ubMDM}{\ubM_{\mathrm{DM}}}
\newcommand{\aMW}{a_{\mathrm{MW}}}
\newcommand{\MLMC}{M_{\mathrm{LMC}}}
\newcommand{\aLMC}{a_{\mathrm{LMC}}}
\newcommand{\rhoLMC}{\rho_{\mathrm{LMC}}}
\newcommand{\sinc}{\mathrm{sinc}}
\newcommand{\ellmax}{\ell_{\max}}

\usepackage{soul} 
\usepackage{xcolor} 
\colorlet{lightmagenta}{magenta!30} 

\colorlet{lightblue}{blue!30}

\colorlet{lightorange}{orange!40}

\definecolor{ao(english)}{rgb}{0.0, 0.5, 0.0}
\colorlet{lightgreen}{ao(english)!40}

\shorttitle{The MW halo linear response to the LMC}
\shortauthors{Rozier et al.}
\graphicspath{{./}{}}

\begin{document}

\title{Constraining the Milky Way halo kinematics via its Linear Response to the Large Magellanic Cloud}

\correspondingauthor{Simon Rozier}
\email{s.rozier@astro.unistra.fr}

\author{Simon Rozier}
\affiliation{Universit\'e de Strasbourg, CNRS, Observatoire astronomique de Strasbourg, UMR 7550, F-67000 Strasbourg, France}
\author{Benoit Famaey}
\affiliation{Universit\'e de Strasbourg, CNRS, Observatoire astronomique de Strasbourg, UMR 7550, F-67000 Strasbourg, France}
\author{Arnaud Siebert}
\affiliation{Universit\'e de Strasbourg, CNRS, Observatoire astronomique de Strasbourg, UMR 7550, F-67000 Strasbourg, France}
\author{Giacomo Monari}
\affiliation{Universit\'e de Strasbourg, CNRS, Observatoire astronomique de Strasbourg, UMR 7550, F-67000 Strasbourg, France}
\author{Christophe Pichon}
\affiliation{CNRS and SU, UMR 7095, Institut d’Astrophysique de Paris, 98 bis Boulevard Arago, F-75014 Paris, France}
\affiliation{IPhT, DRF-INP, UMR 3680, CEA, Orme des Merisiers Bat 774, F-91191 Gif-sur-Yvette, France}
\author{Rodrigo Ibata}
\affiliation{Universit\'e de Strasbourg, CNRS, Observatoire astronomique de Strasbourg, UMR 7550, F-67000 Strasbourg, France}

\begin{abstract}
We model the response of spherical, non-rotating Milky Way (MW) dark matter and stellar halos to the Large Magellanic Cloud (LMC) using the matrix method of linear response theory. Our computations reproduce the main features of the dark halo response from simulations. We show that these features can be well separated by a harmonic decomposition: the large scale over/underdensity in the halo (associated with its reflex motion) corresponds to the $\ell=1$ terms, and the local overdensity to the $\ell\geq2$ multipoles. Moreover, the dark halo response is largely dominated by the first order `forcing' term, with little influence from self-gravity. This makes it difficult to constrain the underlying velocity distribution of the dark halo using the observed response of the stellar halo, but it allows us to investigate the response of stellar halo models with various velocity anisotropies: a tangential (respectively radial) halo produces a shallower (respectively stronger) response. We also show that only the local wake is responsible for these variations, the reflex motion being solely dependent on the MW potential. Therefore, we identify the structure (orientation and winding) of the in-plane quadrupolar ($m=2$) response as a potentially good probe of the stellar halo anisotropy. Finally, our method allows us to tentatively relate the wake strength and shape to resonant effects: the strong radial response could be associated with the inner Lindblad resonance, and the weak tangential one with corotation.
\end{abstract}


\section{Introduction} 
\label{sec:intro}

The nature of \DM\@ is certainly one of the most pressing questions of modern physics. While \DM\@ is generally assumed to consist of a cold and collisionless component of non-baryonic particles, direct evidence for its particle nature is still lacking. While a direct detection might still take a (very) long time, an achievable short-term goal would be to test whether galaxies do indeed reside in \DM\@ halos that are made of a self-gravitating `sea' of collisionless particles, interacting with baryons and with each other through gravity. A distinctive signature of such a self-gravitating sea of particles is that it would react in a predictable way to external perturbations: this response would in principle itself leave a signature in the dynamics of the stellar halos of galaxies, and possibly in their disc dynamics too. Since current and forthcoming large surveys of the \MW\@ are mapping the kinematics of its stellar halo with unprecedented accuracy, predicting and searching for such signatures is timely. Once detected, they could also provide additional information on the \DM\@ distribution in the outskirts of the Galaxy, but also unique information on the underlying {\it phase-space} distribution of \DM\@, which is difficult to get access to otherwise. Hence, studying the response of the \MW\@ \DM\@ halo to external perturbations is in principle a unique way to gain insight both into the existence and nature of \DM\@ and into its phase-space distribution.

In recent years, an array of evidence has emerged to indicate that the main first order perturbation to the \MW\@ halo would come from the infall of the \LMC, whose total mass might represent more than $\sim 10$\% of the \MW\@ mass. Early analyses of its internal dynamics concluded that the mass of the \LMC\@ (with a stellar mass of $2.7 \times 10^9 \, \rm{M}_\odot$) had to be larger than $1.5 \times 10^{10} \, {\rm M}_\odot$, its enclosed mass within 8-9~kpc \citep{Schommer92, vdm2014}. It soon also became clear that the \LMC\@ was most probably on its first infall towards the \MW\@ \citep{Besla2007}, meaning that its \DM\@ halo would be essentially unstripped: combined with the requirement that the \LMC\@ and \SMC\@ have been a long-lived binary, this first infall scenario would imply a total mass of as much as $1.8 \times 10^{11} \, {\rm M}_\odot$ \citep{Kallivayalil2013}. This mass is also well in line with expectations from abundance matching \citep[e.g.,][]{Behroozi2013, Moster2013}. This was later confirmed by a timing constraint integrating backwards the orbits of galaxies currently sitting within 3~Mpc in the Local Volume as well as the relative motion between the \MW\@ and Andromeda, with the \LMC\@ mass as a free parameter, returning a very high mass of $2.5 \times 10^{11} \, {\rm M}_\odot$ \citep{Penarrubia2016}.

The first study of the influence of such a massive \LMC\@ on the dynamics of the Galaxy was conducted by \citet{Gomez2015} who concluded that, due to the displacement of the center of the system, the inner regions of the \MW\@ would be moving with a bulk velocity w.r.t. its outskirts, a reflex motion which would create a dipole in the stellar velocity field. Tentative observational evidence for this reflex motion has been recently provided by \citet{Petersen+2021} and \citet{Erkal2021}. This reflex motion is expected to be accompanied by a local wake trailing behind the \LMC\@ \citep[see, e.g.,][]{Garavito2019}, a phenomenon also tentatively detected by \cite{Conroy2021}. But, as mentioned above, the actual global response of the \DM\@ halo (and, subsequently, of the stellar halo) is more subtle than that, and depends on the underlying distribution of \DM\@ in phase-space, as studied by, e.g., \citet{Laporte2018, Garavito2019, Garavito2021, Tamfal2021}. It has for instance been shown that the LMC could induce an overdensity in the orbital poles of particles moving through the halo \citep{Garavito2021b}, although it has been subsequently shown that the effect was only present for particles with low specific angular momentum \citep{Pawlowski2021}.

Searching for all those signatures obviously requires one to depart from the assumption that the Galaxy is in equilibrium, but starting from equilibrium configurations is still useful as it can serve as a basis for analytic linear perturbation theory, which represents a powerful alternative to numerical simulations. In the present paper, we apply such methods to the specific case of the response of the \MW\@ halo to the infall of the \LMC. They allow us to efficiently cover parameter space, and importantly, to gain physical insight into the various processes and resonances at play, which are obviously more difficult to decipher in $N$-body simulations.

The paper is organised as follows. In Section~\ref{sec:MatrixMethod}, we develop the essential steps of the response matrix formalism, while some analytical details can be found in Appendices~\ref{app:Matrix_method} and~\ref{app:BOB}, some numerical details of the computation in Appendices~\ref{app:Matrix_computation} and~\ref{app:Matrix_operations}, and a validation of the implementation in Appendix~\ref{subsec:ROI_Validation}. Section~\ref{sec:MWResponse} details our models for the \LMC\@ and the \MW, and shows our results for the response of the \MW\@ to the \LMC\@ in a fiducial isotropic case. In Section~\ref{sec:Response_Anisotropic_Halo}, we show that the halo's self-gravity has a minor effect, which prevents us from constraining the \DM\@ phase-space structure from the observation of the stellar halo, but allows us to focus on the forced response of the stellar halo for various anisotropies. In Section~\ref{sec:Discussion}, we discuss the dependence of our results on the details of the \LMC's orbit, and we take full advantage of our method by (i) extracting meaningful information on the \MW's reflex motion and local wake from separating different multipolar components, and by (ii) identifying possible resonant effects acting to build up the wake. We conclude and summarize our results in Section~\ref{sec:Conclusion}.

\section{Linear response theory and the matrix method} 
\label{sec:MatrixMethod}

In order to analytically model the response of the \MW\ to the infall of the \LMC, we resort to the linearisation of the \CBE---Poisson system (or Vlasov-Poisson), a framework known as linear response theory \citep{BinneyTremaine2008}. We first describe here the guiding principles of this analysis, as well as the particular method that we use to tackle this problem, called the matrix method.

\subsection{Linear response theory}

The \MW\ background potential is modelled as a spherical potential $\psi_0(r)$. It is basically representing its \DM\@ halo, which dominates the potential at large radii, but note that it can also include a baryonic component, and that the general method described here is not restricted to spherical potentials. Such potentials are however best suited as a first step for the analysis we conduct hereafter. 

In the absence of any perturber, the orbits of stars and \DM\@ particles in this potential are regular, since their dynamics derives from the spherically symmetric Hamiltonian
\begin{equation}
   H_0(\bx, \bv) = \frac{v^2}{2} + \psi_0(r).
\end{equation}
The orbits are then best described in action-angle phase space coordinates. The actions $\bJ = (J_r, L, L_z)$, where $J_r$ is the radial action, $L$ the norm of the angular momentum and $L_z$ its projection onto the $z$-axis, are then fully labelling regular orbits, while the canonically conjugate angle coordinates indicate where a given particle is located along its orbit. 

According to Jeans' theorem \citep{BinneyTremaine2008}, the phase space \DF\@ of a fully phase-mixed system at equilibrium is only a function of the actions: $F(\bJ)$. Here, we define the equilibrium phase space \DF\@ such that $F(\bx, \bv) \, \rd \bx \rd \bv$ is the {\it mass} located in the phase space volume $\rd \bx \rd \bv$ around $(\bx, \bv)$. As a consequence, we have that $\!\int\! \rd \bx \rd \bv \, F = \Mtot$, the total mass of the distribution of matter, the response of which we choose to study. In other words, the \DF\@ $F$ represents a distribution of orbits (which could be either \DM, stars or both) in the potential $\psi_0$, and linear response theory aims at computing the self-gravitating, collisionless response of this collection of orbits when they are perturbed, in our case by the \LMC's infall. We emphasize that the \DF\@ $F$ needn't self-consistently generate the mean-field potential $\psi_0$, although we will consider that specific case in Section~\ref{subsec:Fiducial_MW_Response}.

The \LMC\@ is modelled as an external perturber with potential $\psi^{\re}$, with a small amplitude compared to the potential $\psi_0$ of the \MW. The \MW's response is represented as a self-induced potential perturbation $\psi^{\rs}$, and is also considered as a source of perturbations on the mean-field orbits, so that the self-gravity of the response is well taken into account. In addition to these extra forces, we consider the reference frame to be accelerated, resulting in a corresponding pseudo-force term in the Hamiltonian. Our accelerated reference frame indeed follows the motion of the \MW\ centre in the asymmetric potential generated by the perturber and the response (the reflex motion, see Section~\ref{sec:intro}). This acceleration corresponds to that of a test particle, initially at rest at the centre of the \MW, and accelerated by the gravitational influence of the total perturbation $\psi_1 = \psi^{\re} + \psi^{\rs}$. All these effects add up to a Hamiltonian $H_0 + \Delta H$ with
\begin{equation}
    \Delta H (\bx, t) = \psi^{\re} (\bx, t) + \psi^{\rs} (\bx, t) + \ba_{\rc}(t) \cdot \bx,
    \label{eq:Perturbing_Hamiltonian}
\end{equation}
where the acceleration of the \MW\@ centre is given by\footnote{Note that this acceleration is different from that used in \cite{Murali1999}. Indeed, the acceleration considered there was that of the host's barycentre. This does not correspond to the acceleration of the host's density centre, which we are following here, because the perturber penetrates the host, creating a shift in the motion of the host's outer parts vs. its inner parts. }
\begin{equation}
    \ba_{\rc}(t) = - \frac{\partial \psi_1(\bx, t)}{\partial \bx}\bigg|_{\bx = \mathbf{0}} = \!\!\int \!\! \rd \bx \, \rho_1(\bx,t) \, \frac{G}{|\bx|^2} \, \mathbf{e_r},
    \label{eq:Centre_acceleration}
\end{equation}
with $\bx$ centred at the \MW\@ centre, $\rho_1 =  \rho^{\re} + \rho^{\rs}$ the density of the perturber and the response, $G$ Newton's constant of gravity, and $\mathbf{e_r} = \bx / | \bx |$. 

In this reference frame, the \CBE\ can be linearised, with the total perturbed distribution function written as $F+f$ with $f \ll F$, giving us the \textit{linearised} \CBE\:
\begin{equation}
    \frac{\partial f}{\partial t} + \bO \cdot \frac{\partial f}{\partial \bT} - \frac{\partial F}{\partial \bJ} \cdot  \frac{\partial \Delta H}{\partial \bT } = 0,
    \label{eq:CBE_linearised}
\end{equation}
where  $\bT = (\theta_1, \theta_2, \theta_3)$ are the angles canonically conjugated to the mean-field actions $\bJ$, and 
\begin{equation}
    \bO = (\Omega_1, \Omega_2, 0) = \frac{\partial H_0}{\partial \bJ}
\end{equation}
are the corresponding mean-field orbital frequencies.

Together with the Poisson equation, $\nabla^2 \psi^{\rs} = 4 \pi G \rho^{\rs}$ with $\rho^{\rs} = \!\!\int\! \rd \bv \, f$, this system of partial differential equations allows for the full integration of our \MW\ model's response to the \LMC.

\subsection{The matrix method}
\label{subsec:Matrix_Method}
To proceed forward, we follow \cite{Kalnajs1977} who undertook a projection of all perturbed quantities onto a bi-orthogonal basis of potentials and densities. This practice gained a renewed interest in recent years \citep[see, e.g.\@,][]{Garavito2021,Sanders2020} for its ability to solve the Poisson equation by construction, allowing for more natural reconstructions of the gravitational potential in simulations. In the context of linear response theory, this technique can also be used to solve the Poisson equation, while the \CBE\@ is transformed into an integral equation in a linear space. 

In short, and as we detail in Appendix~\ref{app:Matrix_method}, we start from a bi-orthogonal basis of potential-density pairs $(\psi^{(p)}\!, \rho^{(p)})$, where $(p)$ typically stands for a triplet of indices $\ell^p, m^p, n^p$ with ${\ell^p \geq 0 }$, ${ |m^p| \leq \ell^p }$, and ${ n^p \geq 0}$, with the bi-orthogonality condition
\begin{equation}
    \!\! \int \!\! \rd \bx \, \psi^{(p)} (\bx) \, \rho^{(q)*} (\bx) = - \delta_p^q,
    \label{eq:biorthcond}
\end{equation}
and we define the projections of the perturbing potentials
\begin{align}
\nonumber
    \psi^{\rs}(\bx, t) & = \sum_{p} a_p(t) \, \psi^{(p)}(\bx), \\
    \psi^{\re}(\bx, t) & = \sum_{p} b_p(t) \, \psi^{(p)}(\bx),
    \label{eq:Projection_perturbing_potentials1}
\end{align}
where the vectors $\ba(t)$ and $\bb(t)$ respectively correspond to the response and external perturbation at time $t$, projected onto the bi-orthogonal basis. 

Then, using the bi-orthogonality condition of eq.~\eqref{eq:biorthcond}, we can write
\begin{equation}
    a_p(t) = - \!\!\int\!\! \rd \bx \!\!\int\!\! \rd \bv \, f(\bx, \bv, t) \, \psi^{(p)*}(\bx),
    \label{eq:ap}
\end{equation}
and we can replace $f(\bx, \bv, t)$ by the solution of the linearized \CBE\@ (eq.~\eqref{eq:CBE_linearised}) assuming that the system is unperturbed at the initial time. Making use of the fact that the integration variables can be canonically changed from $\rd \bx \rd \bv$ to $\rd \bJ \rd \bT$, we then get (see Appendix~\ref{app:Matrix_method})
\begin{equation}
    \ba(t) = \!\!\int_0^t \!\!\! \rd \tau \, \bM(t - \tau) \, \big[\ba(\tau) + \bb(\tau)\big],
    \label{eq:Response_time_integral}
\end{equation}
 where the \textit{response matrix} $\bM(t-\tau)$ contains the full information on the gravitational dynamics in the mean-field equilibrium as a function of time delay $t-\tau$. This matrix is given by
\begin{align}
    \nonumber
    \bM_{pq}(t) \!\! = \!\! - \ri (2 \pi)^3 \! \sum_{\bn} & \!\!\int\!\! \rd \bJ \, \re^{- \ri \, \bn \cdot \bO \, t} \, \bn \!\cdot\! \frac{\partial F}{\partial \bJ} \\ 
    & \times \psi_{\bn}^{(p)*\!}(\bJ) \, \big(\psi_{\bn}^{(q)\!}(\bJ) \!+\!  \phi_{\bn}^{(q)\!}(\bJ) \big).
    \label{eq:Response_Matrix}
\end{align}
Some elements of this expression derive from the Fourier transform of spatial functions w.r.t. the angles (see eq.~\eqref{eq:Fourier_transform_angles}), the resonance vector of integers $\bn = (n_1, n_2, n_3)$ labelling the corresponding Fourier coefficients. In more detail, $\psi_{\bn}^{(p)}$ results from the transformation of the potential basis function with index $(p)$, and $\phi_{\bn}^{(q)}$ emerges from the transformation of the $(q)$ component of the non-inertial fictitious potential, given by
\begin{equation}
    \phi^{(q)}(\bx) = \bx \cdot \!\! \int \!\! \frac{G \, \rd \bx}{|\bx|^2} \, \rho^{(q)}(\bx) \, \mathbf{e_r}.
    \label{eq:Definition_inertial_term}
\end{equation}


The details of the demonstration of eqs.~\eqref{eq:Response_time_integral} and~\eqref{eq:Response_Matrix} are given in Appendix~\ref{app:Matrix_method}, while the particular choice of the bi-orthogonal basis for the present study \citep{CluttonBrock1973} is detailed in Appendix~\ref{app:BOB}. 

Equation~\eqref{eq:Response_time_integral} highlights how this method incorporates the response's self-gravity. Indeed, not only does the response $\ba(t)$ derive from the influence of the perturber at all time steps $\bb$, via the dynamics encoded in the response matrix, but the response at the current time also derives from the influence of the response itself at all previous time steps, as represented by the term $\ba(\tau)$ in the r.h.s. If self-gravity could be neglected, one would only need to replace $\ba(\tau)$ by $\mathbf 0$ to compute the system's response to the perturber (see Section~\ref{subsec:Bare_Response}).

While the matrix method was mostly used in its ``frequency'' version (after a Laplace transform of all time-varying quantities) to detect linear instabilities in various types of self-gravitating systems \citep[see, e.g.,][]{Zang1976,Polyachenko1981,Weinberg1991,Vauterin1996,DeRijcke+2019,Breen2021}, the present ``time'' version has scarcely been used, despite its ability to solve for the self-gravitating response of perturbed stellar systems (but see \cite{Seguin1994,Weinberg1998,Murali1999}, and \cite{Pichon2006} for noticeable, if not unique, exceptions). We therefore hope that the present study will help reviving the interest of the community in this approach.

\subsection{From time integration to matrix inversion}

In the form of eq.~\eqref{eq:Response_time_integral}, the problem is not yet explicitly linear, in the sense that there remains a step to directly relate the perturber $\bb$ and the response $\ba$ through a linear relation. Indeed, it appears that the response at the current time, $\ba(t)$, explicitly depends on the self-induced perturbation at all previous times, as given by the $\ba(\tau)$ term in the integrand of the r.h.s. To do so, we will approximate the time integral by its Riemann sum, therefore exhibiting the intrinsic linearity of the problem.

Let us assume that we aim at computing the linear response for a full period of time $[0, T]$. Let us next divide this time interval in $K + 1$ steps $0 = t_0 < \dots < t_K = T$. At each step $i \geq 1$, eq.~\eqref{eq:Response_time_integral} can be approximated as
\begin{equation}
    \ba(t_i) = \sum_{j=0}^{i-1} \Delta t \, \bM(t_i - t_j) \, \big[\ba(t_j) + \bb(t_j)\big],
    \label{eq:Discrete_integration}
\end{equation}
using the rectangular rule at the lower bound of each step, with $\Delta t = T/K$. Let us now define $\uba$ (resp. $\ubb$) as the vector built by stacking all vectors $\ba(t_0), ..., \ba(t_K)$ (resp. $\bb(t_0), ..., \bb(t_K)$) on top of each other. Furthermore, the matrix $\ubM$ is defined by blocks, so that the block in the line $i$ and column $j$ is given by
\begin{equation}
\ubM_{ij} = \left\{\begin{aligned}
&\Delta t \, \bM(t_i - t_j) & \, \mathrm{for} \, j < i, \\
& \mathbf{0} & \, \mathrm{for} \, j \geq i.
\end{aligned}\right.
\label{eq:Full_time_matrix}
\end{equation}
Here, the vectors $\uba$ and $\ubb$ contain the information on the external and induced perturbations over the full time interval $[0, T]$, and the matrix $\ubM$ contains the information on the system's linear dynamics over all possible time delays. With these definitions, eq.~\eqref{eq:Discrete_integration} can be rewritten as a matrix product as
\begin{equation}
\uba = \ubM \, (\uba + \ubb).
\label{eq:Response_Nonlinear_Form}
\end{equation}
According to eq.~\eqref{eq:Full_time_matrix}, the matrix $\ubI - \ubM$ can always be inverted, where $\ubI$ is the identity matrix of suitable size, and the system's response over the full time interval can be computed by a simple product of matrices through
\begin{equation}
\uba = [\ubI - \ubM]^{-1} \, \ubM \, \ubb = \big( [\ubI - \ubM]^{-1} \!\! - \ubI \big) \, \ubb.
\label{eq:Linear_response_final}
\end{equation}
In this form, the problem is evidently linear, and its resolution can be summarised in the following steps: (i) take a model for the external perturber's density at each time step, and project it onto the bi-orthogonal basis to get the full perturbing vector $\ubb$; (ii) compute the full response matrix $\ubM$ following eqs.~\eqref{eq:Response_Matrix} and~\eqref{eq:Full_time_matrix}; (iii) perform the matrix inversion and multiplication according to eq.~\eqref{eq:Linear_response_final} to compute the host system's response $\uba$ at each time step. These are the steps we take hereafter to compute the \MW's response to the \LMC, as detailed in the next section.

One asset of response theory is to provide an explicit linear relationship via eq.~\eqref{eq:Linear_response_final} between the response of the \MW\@ halo, a parametric representation of the underlying property of the unperturbed equilibrium, and  the properties of the perturbation. While observing the former and the latter, one can constrain the corresponding parameter, hence e.g.\@ probe the internal kinematics of the \MW\@ halo.

\subsection{Two-component system}
\label{subsec:Two_Component_Matrix}

In order to describe a \MW\@ model made of stars and \DM, let us introduce the linear response of a system made of two sub-components. For that purpose, we follow the work of \cite{Weinberg1998} in this section.

Let us consider the \MW\@ halo to be made of a dominant \DM\@ component of mass $\MDM$, and a light stellar component of mass $M_*$, so that $\MMW = \MDM + M_*$ and $M_* / \MMW = \epsilon \ll 1$. Note that we neglect here the stellar disk, which we effectively absorb within the DM component, so that $M_*$ only represents the stellar halo, with $\epsilon \sim 10^{-3}$. Let us assume that both components follow the identical density profile, which adds up to the total potential $\psi_0$, and that they are described by the \DFs\@ $\FDM$ and $F_*$. We can therefore associate the response matrices $\ubMDM$ and $\ubM_*$ (constructed from eqs.~\eqref{eq:Response_Matrix} and~\eqref{eq:Full_time_matrix}) to each of those components, computed using the same basis elements. In \cite{Weinberg1998}, we learn that the generalisation of our eq.~\eqref{eq:Response_Nonlinear_Form} is given by
\begin{align}
\nonumber
    \begin{pmatrix}
    \ubaDM \\
    \uba_* 
    \end{pmatrix}
    =
    \begin{pmatrix}
    \ubMDM & \ubMDM \\
    \ubM_* & \ubM_*
    \end{pmatrix}
    & \cdot
    \begin{pmatrix}
    \ubaDM \\
    \uba_* 
    \end{pmatrix} \\
    +
    \begin{pmatrix}
    \ubMDM & \mathbf{0} \\
    \mathbf{0} & \ubM_*
    \end{pmatrix}
    & \cdot
    \begin{pmatrix}
    \ubb \\
    \ubb 
    \end{pmatrix},
\end{align}
where $\ubaDM$ and $\uba_*$ are the responses in each sub-component, and the same perturber $\ubb$ is applied. The respective responses of the two sub-components therefore verify the system of coupled equations
\begin{subequations}
\begin{equation}
\ubaDM = \ubMDM \, (\ubaDM + \uba_* + \ubb),
\end{equation}
\vspace{-10pt}
\begin{equation}
\uba_* = \ubM_* \, (\ubaDM + \uba_* + \ubb).
\end{equation}
\end{subequations}
Since we assumed $\epsilon \ll 1$, then the response of the stellar halo is generically negligible in mass w.r.t. that of the \DM, i.e. $\uba_* \ll \ubaDM$, and the system of equations becomes
\begin{subequations}
\begin{equation}
\ubaDM = ([\ubI - \ubMDM]^{-1} - \ubI) \, \ubb,
\end{equation}
\vspace{-12pt}
\begin{equation}
\uba_* = \ubM_* \, [\ubI - \ubMDM]^{-1} \, \ubb.
\label{eq:Stellar_Response_Two_Components}
\end{equation}
\end{subequations}
This last equation reveals how the kinematics of the \DM\@ halo can impact the response of the stellar halo. 
Indeed, 
if $[\ubI - \ubMDM]^{-1}$ is significantly different from the identity matrix, 
then the stellar response depends on the kinematic state of the \DM\@ via its response matrix $\ubMDM$. As we will see in Section~\ref{subsec:Bare_Response}, $[\ubI - \ubMDM]^{-1}$ describes how much the \DM's self-gravity impacts the structure of the response: the larger the impact of self-gravity, the further from identity $[\ubI - \ubMDM]^{-1}$ is.

\section{The \MW's response to the \LMC} 
\label{sec:MWResponse}

In this section, we describe our model for the \MW\@ and the \LMC, we apply the matrix formalism to that interaction, and analyse our results.

\subsection{Models for the \MW\@ and the \LMC}
\label{subsec:Models_MW_LMC}

In order to qualitatively compare our results to those of $N$-body simulations from the literature, we chose to represent the \MW\@ and the \LMC\@ with models that are resembling those of the fiducial simulation of \cite{Garavito2021}. In the present case, the \MW\@ is fully described as a self-consistent spherical halo, incorporating both \DM\@ and stellar halo components. It is modelled as a Hernquist sphere with a  \DF\@ from \cite{Baes2007} (see Appendix~\ref{app:Baes} for details), which self-consistently generates its total potential. The total \MW\@ mass is taken to be $\MMW = 1.57 \times 10^{12} M_{\odot}$, and the scale radius of the Hernquist profile is ${\aMW = 40.8 \, \mathrm{kpc}}$. The halo is isotropic, i.e. it has a constant $\beta$ parameter set to 0. These characteristics define the mean-field potential ($\psi_0$) and phase space \DF\@, $F(E,L)$, which are used in the computation of the response matrix (eq.~\ref{eq:Response_Matrix}). Note that this setup straightforwardly enters the two-component description of Section~\ref{subsec:Two_Component_Matrix} by considering that the \DFs\@ of the \DM\@ and the stars are proportional, with $F_* = \epsilon \, F$ and $\FDM = (1-\epsilon) \, F$, with $\epsilon \sim 10^{-3}$. In that case, the response matrices and the responses are also proportional, with the same relations between the \DM\@ and stellar quantities.

The \LMC\@ is also modelled as a Hernquist sphere, with a total mass of $\MLMC = 1.8 \times 10^{11} M_{\odot}$, and a Hernquist scale radius of $\aLMC = 20 \, \mathrm{kpc}$. Note that only the \LMC\@ density matters here, not its internal dynamics, as it is merely considered as a gravitational perturber to the \MW. This density is used to construct the vector $\ubb$.

\begin{figure}
    \centering
    \includegraphics[width=0.47\textwidth]{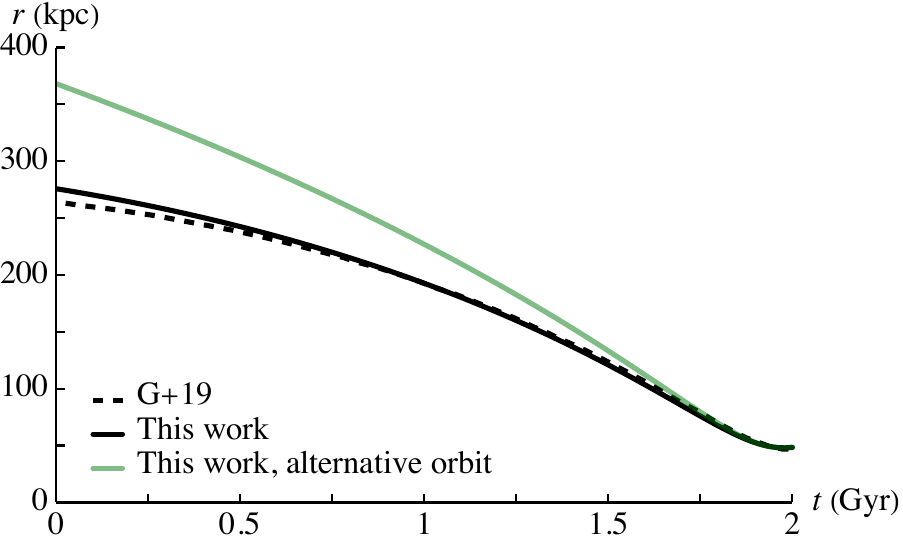}
    \caption{Evolution of the distance from the \MW\@ centre to the \LMC\@ centre, as given (dashed) by \cite{Garavito2019}, (full, black) from a leap-frog integration in a static Hernquist \MW\@ potential, and (green) from a more realistic setup described in Section~\ref{subsec:Influence_LMC_Orbit}. Our fiducial orbit in black is relatively close to that of \cite{Garavito2019} in terms of the evolution of the \LMC's distance to the \MW, which does not exclude differences in the azimuthal path of the \LMC.} 
    \label{fig:LMC_orbit_distance}
\end{figure}

In order to represent the \LMC's infall onto the \MW, we simply integrated the orbit of a particle in the aforementioned \MW\@ potential, starting from estimates of the position and velocity of the \LMC\@ at its pericentre ($\sim 50 \,\mathrm{Myr}$ ago): $r_{\mathrm{p,LMC}} = 48 \, \mathrm{kpc}$, $v_{\mathrm{p,LMC}} = 340 \, \mathrm{km.s}^{-1}$ \citep{Salem+2015}. We integrated this orbit using a leap-frog algorithm with 100 time steps per $\mathrm{Gyr}$, and selected the portion of the orbit which covers the infall of the \LMC\@ since $2 \, \mathrm{Gyr}$ ago. The distance of the \LMC\@ to the \MW\@ centre is represented as a function of time in Fig.~\ref{fig:LMC_orbit_distance}. It may seem that this trajectory is very close from that of the \LMC\@ in the simulations from \cite{Garavito2019}, however in our case, there is no shift between the \MW\@ \COM\@ and its cusp in the construction of the orbit, while they represent the distance to the \MW\@ \COM. Additional differences in the shape of the \MW\@ potential also result in differences in the shape of the \LMC's orbit in its plane. We keep this simple orbit as our fiducial setup, and discuss the possible influence of the \MW's motion later in Section~\ref{subsec:Influence_LMC_Orbit}. In that discussion, we build the orbit represented in green in Fig.~\ref{fig:LMC_orbit_distance}. From our fiducial trajectory, we extracted $K + 1=21$ equally spaced time steps to represent the \LMC's orbit, i.e.\@ $\Delta t = 100 \, \mathrm{Myr}$. We finally have a succession of mass density profiles for the \LMC\@ in the frame centered on the \MW\@ centre, $\rhoLMC(\bx, t_i)$ for $0 \leq i \leq K$.

This time interval may seem large for $N$-body simulations, however it is appropriate for the linear theory. Indeed, in the case of $N$-body simulations, the inertial motion of a particle between two time steps is a straight line of constant velocity, as if the particle was isolated. This implies large discrepancies in the particles' orbits when the time resolution is not high enough. For the response matrix method, three key features still occur between two time steps: (i) the system's response still follows the orbits in the mean-field potential, (ii) the system is still responding to the perturber, as if it had not moved from the previous time step, and (iii) the system is still responding to the response itself (the response is self-gravitating), as if it had not moved since the previous time step. These features imply that there is a much lower amount of time steps required to compute the system's response with a reasonable accuracy. Typically, the matrix method requires a rough representation of the positions that the perturber takes in its motion, while $N$-body simulations require a fine integration of the orbits in the host.

\subsection{Projection of the \LMC\@ onto the basis}

\begin{figure*}
\centering
    \begin{minipage}{0.77\textwidth}
    \centering
    \includegraphics[width=1.0\textwidth]{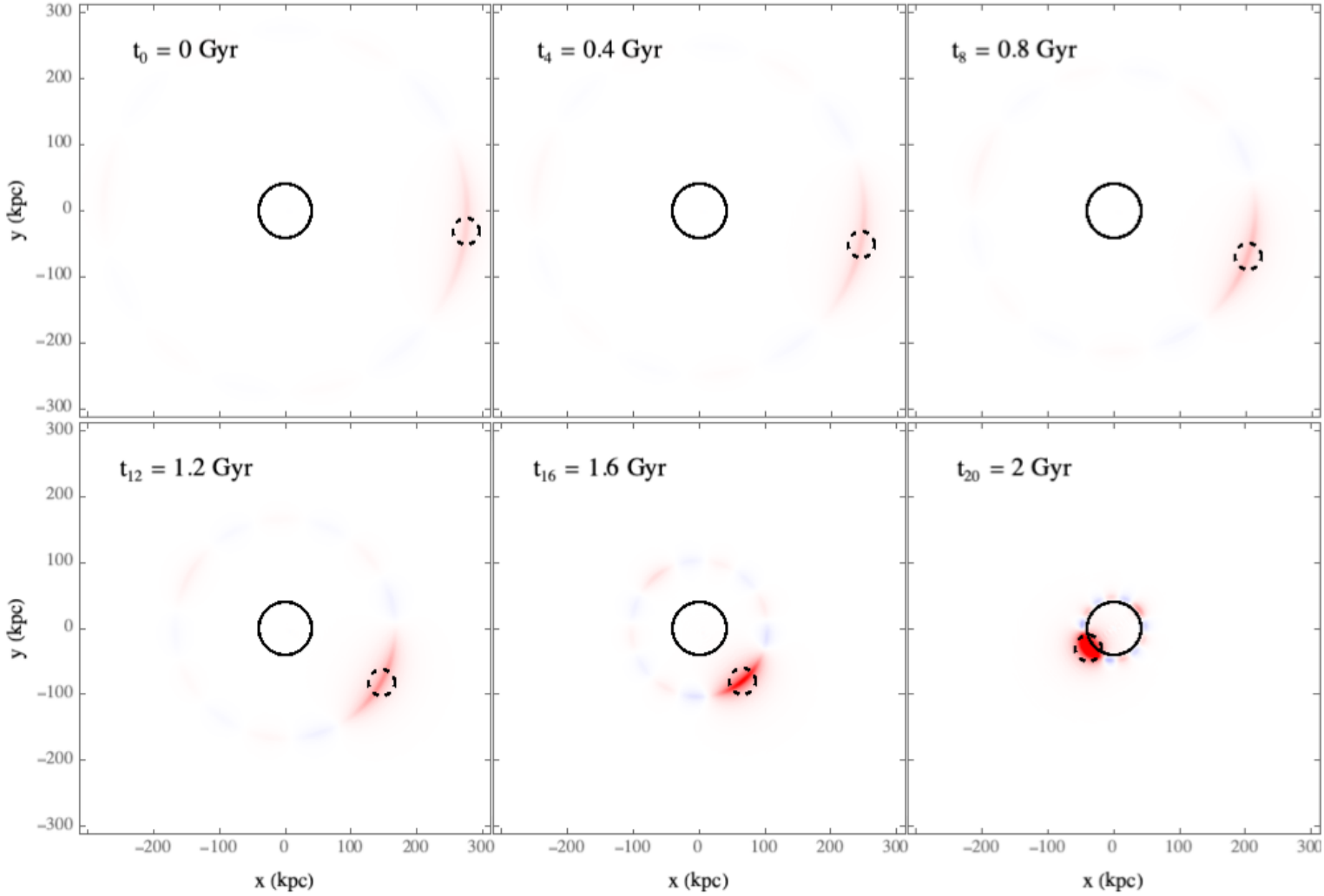}
    \end{minipage}
    \begin{minipage}{0.21\textwidth}
    \subfloat{
    \centering
    \includegraphics[width=0.95\textwidth]{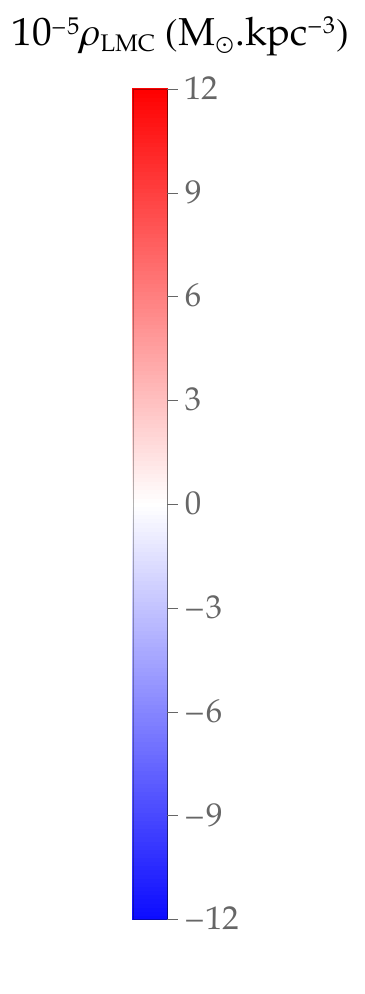}}
    \end{minipage}
    \caption{Reconstruction of the \LMC\@ density after a projection onto the bi-orthogonal basis. $t=0 \, \mathrm{Gyr}$ corresponds to the initial conditions, $2 \, \mathrm{Gyr}$ ago. The full circle represents the \MW\@ scale radius $\aMW$, while the dotted circle is the \LMC\@ scale radius, $\aLMC$. Although it has a limited angular resolution, the basis reconstruction marks well the position of the \LMC\@ at all time steps. The reconstruction is more accurate at late times, when the \LMC\@ is close to the \MW. The mass of the \LMC\@ is reconstructed by the $\ell = 0$ component, therefore it is unaffected by the limited angular resolution.}
\label{fig:LMC_density_evolution}
\end{figure*}

In order to apply the matrix formalism to the \MW-\LMC\@ interaction, we need to project the perturber (the \LMC) onto the bi-orthogonal basis, to compute the perturbing vector $\ubb$. To do so, we simply apply eq.~\eqref{eq:Projection_coefficient} with the density of the \LMC\@ (considering the \LMC's orbital plane as the $z=0$ plane) and the basis defined by eq.~\eqref{eq:Basis_general_form} and Appendix~\ref{app:BOB}. This gives
\begin{equation}
    b_p(t_i) = - \!\! \int \!\! \rd \bx \, \rhoLMC(\bx, t_i) \, \psi^{(p)}(\bx).
    \label{eq:Projection_perturber}
\end{equation}
Recall that the index $(p)$ actually stands for the three integers $m^p, \ell^{p}, n^p$. In our basis, $m^p$ and $\ell^p$ identify the angular harmonic order of the basis element, while $n^p$ identifies its radial order. Applying the projection of eq.~\eqref{eq:Projection_perturber} to a series of $(p)$ yields the sub-vector $\bb(t_i)$, which is then computed at all times $t_0 < \cdots < t_K$ to give the stacked vector $\ubb$.


In theory, the potential-density basis has an infinite number of elementary functions, so that an infinite number of projection coefficients is required to reconstruct the perturber with arbitrary precision. In practice, the basis should be truncated in both its angular and radial orders. These orders of truncation are parameters that should be tested for convergence to ensure the robustness of our results. In the angular direction, our fiducial choice is $\ellmax = 6$, while the definition of the spherical harmonics always imposes that $-\ell \leq m \leq \ell$. This choice is motivated by two facts: first, \cite{Garavito2021} show that most of the information on the \MW's response is contained within low harmonics $\ell \leq 4$. Second, as is shown in Appendix~\ref{subsec:Spherical_matrix}, in a spherical, non-rotating system, there is no coupling in the system's response between different angular harmonics. Therefore, reconstructing the \MW's response up to $\ell = 6$ only requires us to project the \LMC\ up to the same harmonic order. For each of these harmonics, we restrict $m$ to values such that $0 \leq m \leq \ell$ and $\ell - m$ is even, because we are dealing with real fields which are symmetric w.r.t. the equatorial plane. In terms of radial truncation, we found that using a fiducial maximal order of $\nmax = 200$ represented a good balance between the accuracy of the reconstruction and the computational time. We therefore have a total of 3216 basis functions.

Figure~\ref{fig:LMC_density_evolution} shows how the \LMC's density in its orbital plane is reconstructed once it has been projected onto this truncated basis. Strikingly, the quality of the spatial reconstruction of the \LMC\@ is not constant through time. In particular, the \LMC\@ appears as a shallow angular ripple at the beginning of its infall, when it is the furthest from the \MW\@ centre. This can be mainly explained by the low value of $\ellmax$. Indeed, the angular resolution of the basis is approximately given by $\pi / \ellmax$, so that when the \LMC\@ is far from the \MW\@ centre, it is too small to be well resolved. We checked, however, that the \LMC's total mass is well recovered, even in the first time steps\footnote{Indeed, the information of the \LMC\@ mass is only borne by the $\ell = 0$ harmonics, so that the quality of the mass reconstruction only depends on the radial truncation.}. By the end of the interaction, the \LMC\@ is much closer to the \MW\@ and its angular structure can be better resolved by our basis. In terms of radial reconstruction, it seems that the position and structure of the \LMC\@ are well reconstructed after the projection. This is expected, as we have used a relatively large number of radial elements.

\begin{figure*}
\centering
    \begin{minipage}{0.87\textwidth}
    \centering
    \includegraphics[width=1.0\textwidth]{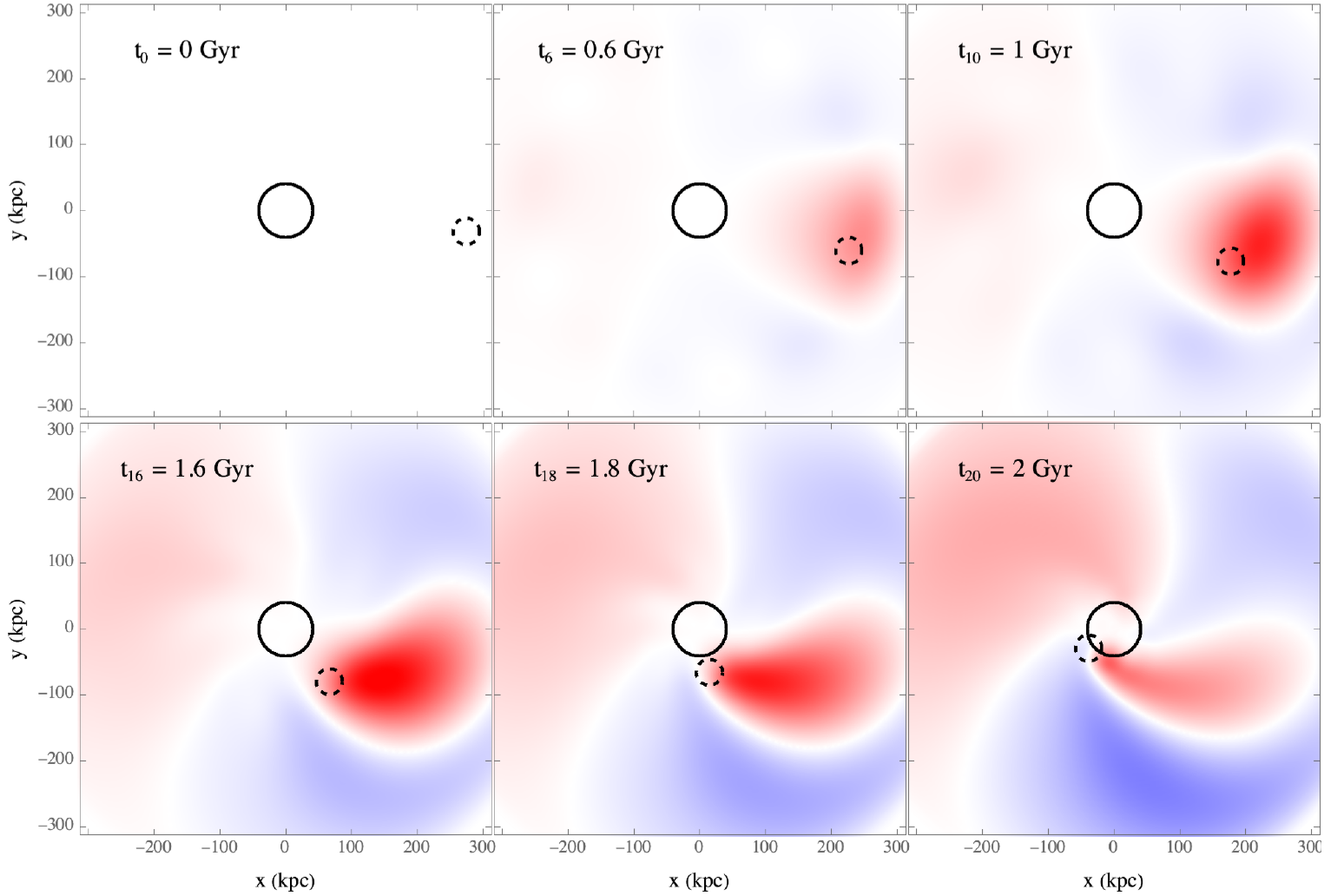}
    \end{minipage}
    \begin{minipage}{0.1\textwidth}
    \subfloat{
    \centering
    \includegraphics[width=0.95\textwidth]{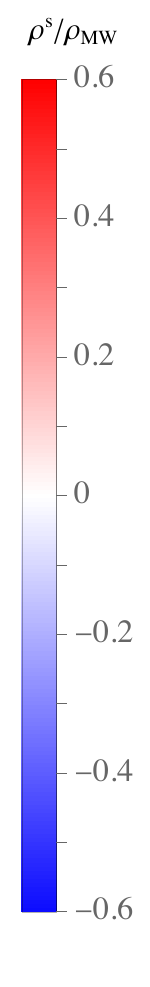}}
    \end{minipage}
    \caption{Reconstruction of the response of our fiducial isotropic \MW\@ DM\@ + stellar halo in terms of relative over-density in the \LMC’s orbital plane.}
\label{fig:MW_response_evolution}
\end{figure*}
        
\subsection{Results: response of the \MW}
\label{subsec:Fiducial_MW_Response}

Once the perturbing vector $\ubb$ is computed, the response is recovered by applying eq.~\eqref{eq:Linear_response_final}. The details of the computation of the response matrix for a spherical non-rotating system are given in Appendix~\ref{app:Matrix_computation}, and we used tailor-made matrix inversion and multiplication algorithms as described in Appendix~\ref{app:Matrix_operations}. In particular, the response matrix is given by eqs.~(\ref{eq:Response_matrix_spherical}-\ref{eq:Definition_function_P}) as an integral over the two actions $\tbJ = (J_r,L)$, and as a sum over the corresponding ``in-plane" resonance vectors $\tbn=(n_1, n_2)$. We checked for the convergence of the following results w.r.t. our many computational parameters, in particular, we observed little variation when we doubled or halved the basis scale radius $\Rb=11\, \aMW$, or when we increased the maximum radial Fourier number $n_{1 \max} = 2$. Convergence w.r.t. the number of radial basis elements is also comfortably reached.


Figure~\ref{fig:MW_response_evolution} shows the response of the \MW\@ along the \LMC's infall, in terms of the relative overdensity w.r.t.\@ the \MW\@ background density, $\rho^{\rs}/\rho_0$, in the \LMC's orbital plane. This figure, and particularly its last panel, is consistent with the corresponding figure in \cite{Garavito2021}.  This consistency indicates that linear response theory is able to realistically reproduce the self-consistent response of the \DM\@ and stellar halos to the \LMC's infall. 

In Fig.~\ref{fig:MW_response_evolution}, we can identify and follow two particular features in these density maps: on the one hand, there is an overdensity which trails behind the \LMC's trajectory. This feature emerges as a spatially large overdensity in the first $\sim 1.6 \, \mathrm{Gyr}$ of the interaction, when the \LMC\@ is slow and can attract some of the \MW\@ in its vicinity. Later on, the \LMC\@ starts falling faster towards the \MW\@ centre, and the overdensity starts moving behind the \LMC's motion, and decays because the \LMC\@ does not stay long enough to attract more material at a single place. Since this overdensity quickly disappears as the \LMC\@ moves away, it is identified as the transient response of the \MW\@ \citep[see, e.g.,][]{Garavito2021}, also called the \LMC's wake. On the other hand, we can identify a large scale dipolar over/underdensity oriented in the direction of the \LMC\@ (with the overdensity opposing the \LMC), which grows in amplitude all along the \LMC's infall. This feature can be associated with the shift in the dynamics of the \MW\@ centre w.r.t.\@ that of its outer parts, i.e. the MW's reflex motion. Since our reference frame follows the \MW\@ cusp, parts of the \MW\@ which are further away than the \LMC\@ get shifted w.r.t. the current reference frame at each time step. This explains (i) that the dipolar feature grows in amplitude all along the interaction, as the \LMC\@ roughly stays on the same side of the \MW, and (ii) that the dipolar over/underdensity populates more central regions of the \MW\@ as the \LMC\@ gets closer to the centre. We will discuss these features in more detail in Section~\ref{sec:Discussion}.

\section{Anisotropic stellar halo} 
\label{sec:Response_Anisotropic_Halo}

Now that we have shown that the matrix method is able to quantitatively compute the \MW\@ halo response to the \LMC, we can take advantage of its numerical efficiency to repeat the experiment on different halo models with varying velocity anisotropy. In strongly anisotropic spheres, it was shown \citep[see, e.g.,][]{Henon1973,Palmer1989,Rozier2019} that linear instabilities can occur, which makes these structures unrealistic. However, we will first show that in our fiducial model there is little impact of the \MW's self-gravity on the response, in agreement with the results of \cite{Seguin1994} in a similar setup. This implies that we can treat the response of the stellar halo separately, since the absence of self-gravity means that the stellar halo is essentially insensitive to the response of the \DM\@ halo. We will therefore show the forced response of models of the stellar halo with various velocity anisotropies.

\subsection{The influence of self-gravity}
\label{subsec:Bare_Response}

As we previously highlighted in Section~\ref{subsec:Matrix_Method}, the matrix method straightforwardly allows for the measurement of the response's self-gravity. In practice, we can compare the self-gravitating response $\ubas$, as obtained from eq.~\eqref{eq:Linear_response_final}, to the bare response $\ubab$, defined by
\begin{equation}
    \ubab = \ubM \, \ubb.
    \label{eq:Bare_Response}
\end{equation}
The difference between these two responses resides in the inclusion or not of the response $\uba$ in the r.h.s. of eq.~\eqref{eq:Response_time_integral}. Formally, they satisfy the relation
\begin{equation}
    \ubas = \ubab + \sum_{i=2}^{\infty} \ubM^i \, \ubb.
\end{equation}
The bare response only includes the response to the perturber at the first gravitational order (i.e., a single application of the response matrix). The self-gravitating one further includes the response induced in the system by this first order response (i.e., $\ubM (\ubM \, \ubb)$), and recursively at all other orders (i.e., all $\ubM^i \, \ubb$ with $i \geq 3$). Alternatively (see eq.~\eqref{eq:Linear_response_final}), $\ubas$ and $\ubab$ are also related through
\begin{equation}
    \ubas = [\ubI - \ubM]^{-1} \, \ubab.
\end{equation}
It shows that the influence of self-gravity is represented by how much $[\ubI - \ubM]^{-1}$ deviates from the identity matrix.

Computing the bare response follows similar steps as the self-gravitating one, as described in Appendix~\ref{app:Matrix_operations}. Figure~\ref{fig:MW_Bare_Residuals} shows the bare response of the \MW\@ to our \LMC\@ model at the last time step, as well as the residuals when it is subtracted to the self-gravitating one. Although it is interesting to see that the region where self-gravity seems most active is within the small scale wake, it is striking that the bare response is the dominant component in the full self-gravitating response, while higher order loops have smaller contribution. 

This has very important consequences on the way this problem can be analysed with the linear theory, as well as on the hope for constraining the phase-space structure of the \MW's \DM\@ halo. If we want to constrain the \DM\@ halo's response, we would need this response to influence an observable population of stars, such as the stars in the stellar halo. Coming back to our model of the \MW\@ as a two-component system (Section~\ref{subsec:Two_Component_Matrix}), with a stellar and a \DM\@ halo, what we have shown is that the total response verifies $[\ubI - \ubM]^{-1} \simeq \ubI$. If we exclude the possibility that the \DM\@ and stellar components conspire into such a result, it means that the influence of the \DM's self-gravity is low. Complementary results additionally show that this conclusion is still true if the \DM\@ halo is anisotropic. Following eq.~\eqref{eq:Stellar_Response_Two_Components}, this implies that the \DM\@ kinematic state has little influence on the stellar halo response. The latter therefore mainly corresponds to the forced impact of the perturber ($\uba_* \simeq \ubM_* \, \ubb$). 
As a consequence, there is little hope for strongly constraining the kinematics of the \DM\@ halo from the response of the \MW\@ to the \LMC.




Let us emphasize that the minor influence of self-gravity should not be regarded as a property of the \MW, but as a consequence of the merger's young age.
Indeed, while the ability to amplify perturbations through self-gravity is a property of the initial equilibrium, the self-gravitating wakes still take time to build up, even in a highly responsive system. 
Formally, the response matrix bears the information on the halo's dynamics, 
but only for the finite time during which we are modelling it. When the integration time $T$ is small, the matrix $[\ubI - \ubM]^{-1}$ cannot get very far from the identity, whatever the \MW's kinematic state, and only when $T$ grows can the matrix significantly deviate from identity, and this deviation should depend on the halo's kinematics. 
In this work, the \LMC\@ is on its first infall, so that the interaction is too short for such self-gravitating wakes to develop, even when the \MW\@ halo is assumed anisotropic.
In a different setup where a satellite is on a periodic orbit around the galaxy for a long time (an infinite time, really), \cite{Weinberg1989} shows that self-gravity has a significant influence on the galaxy's response, even in an isotropic case. 

\begin{figure}
\centering
    \begin{minipage}{0.48\textwidth}
    \centering
    \includegraphics[width=0.80\textwidth]{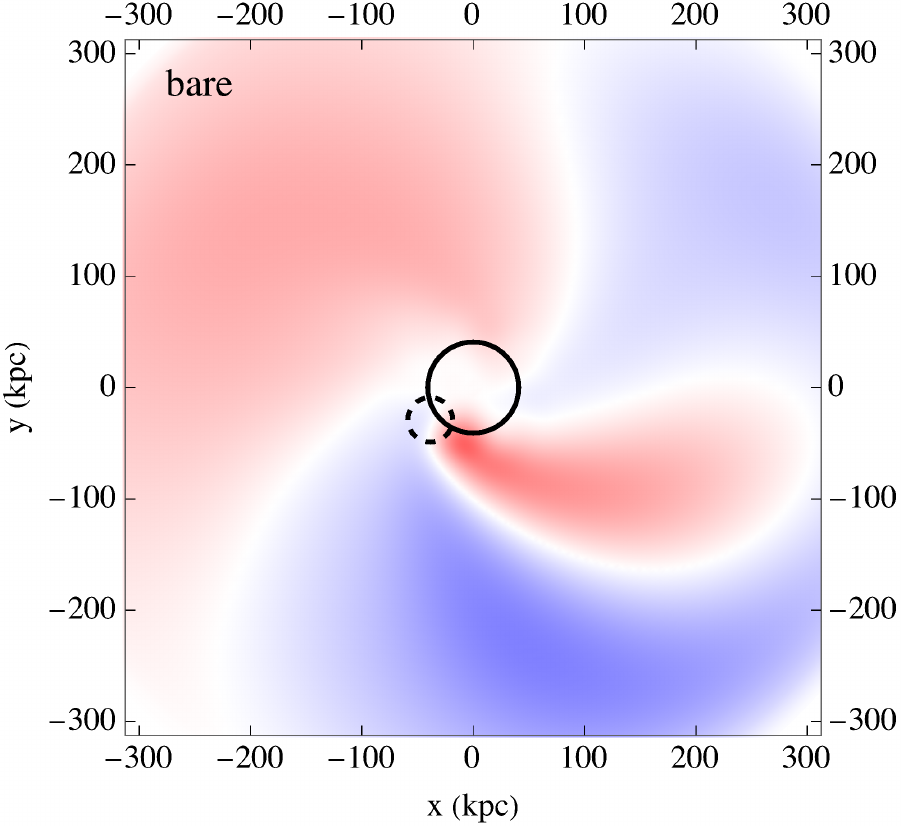} 
    \hspace{-0.12cm}
    \includegraphics[width=0.135\textwidth]{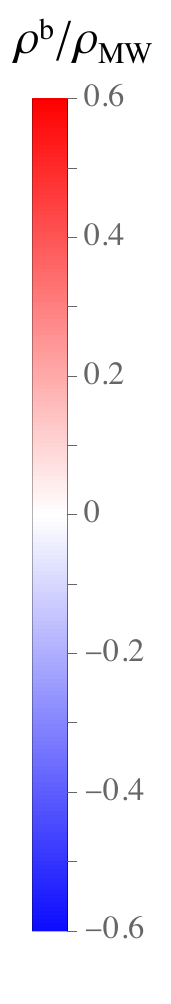}
    \hspace{0.22cm}
    \end{minipage}
    \begin{minipage}{0.48\textwidth}
    \centering
    \includegraphics[width=0.80\textwidth]{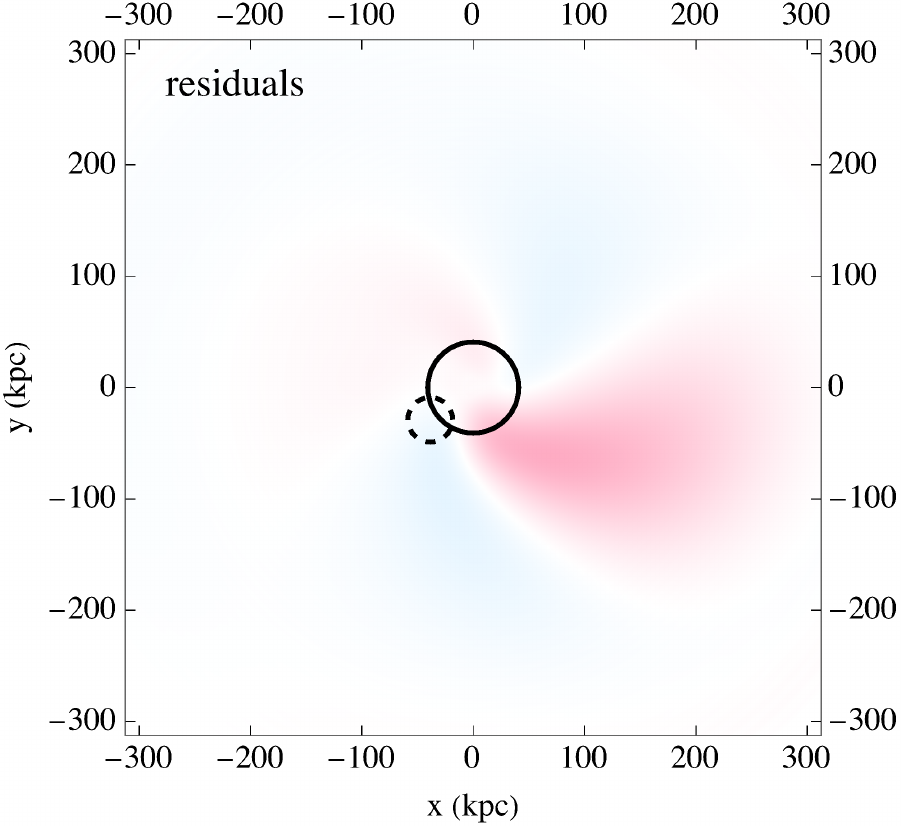} 
    \hspace{-0.4cm}
    \includegraphics[width=0.21\textwidth]{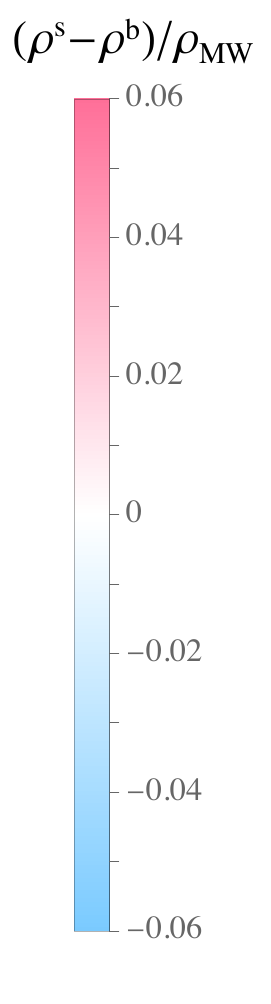}
    \end{minipage}
    \caption{Bare \MW\@ response (top panel) and residuals when compared with the self-gravitating one (bottom panel). The scale of the residuals is about 10 times smaller than the bare response. Self-gravity is therefore responsible for less than 10\% of the total response. }
\label{fig:MW_Bare_Residuals}
\end{figure}

\submitjournal{in the end, check that this is still true (in particular the ref to where we do this)}
The fact that it is possible to neglect the effect of self-gravity in the \MW's response also has positive consequences for our analysis. Indeed, it means that the response of the \MW\@ stellar halo to the \LMC\@ is essentially independent from the phase space distribution of the \MW\@ \DM\@ halo. We can therefore modify our approach, and focus on the bare response of the stellar halo only, considering the \DM\@ halo as an inert gravitational component. We perform this study for stellar halos with different orbital anisotropies in Section~\ref{subsec:Stellar_Halo_Response} below, to evaluate how the strength of the response of the stellar halo depends on its underlying phase-space distribution.

\subsection{Response of stellar halos with different anisotropies}
\label{subsec:Stellar_Halo_Response}

Starting here, and until the end of Section~\ref{sec:Discussion} (except briefly in Section~\ref{subsec:Influence_LMC_Orbit}), we shift from studying the full self-gravitating response of the \MW\@ \DM\@ and stellar halos altogether to studying the bare response of the stellar halo only. Formally, we are assuming that the response of the stellar halo is not influenced by the \DM\@ halo's self-gravity, i.e. we are replacing $[\ubI - \ubMDM]^{-1}$ by $\ubI$ in eq.~\eqref{eq:Stellar_Response_Two_Components}. For that purpose, we take the \DM\@ halo as a background, inert potential, and we consider the stellar halo as a low mass component responding in the potential of the \MW's \DM\@ halo, but with its own density and kinematic structure. More precisely, the background potential is still the same Hernquist sphere, but now the phase space \DF\@ $F_*$ only represents the stellar halo, normalised with a total mass $\Mtot = M_* = \epsilon \, \MMW$ (with $\epsilon \ll 1$). We consider that the stellar halo is distributed according to a Hernquist density, denoted $\rho_*$, with a \DF\@ also given by eq.~\eqref{eq:Baes_DF}, but rescaled by the factor $\epsilon$. Note that the total mass of the stellar halo need not be specified, since (i) we consider the bare response of the stellar halo, which has a linear dependence in its mass, and (ii) all quantities we consider are relative to the initial stellar density $\rho_*$, so the linear dependence w.r.t. $M_*$ is dropped.

Now, we let the stellar halo have a different kinematic structure by changing the value of $\beta$: due to their different formation scenarios, we can expect the stars to present different kinematics from the \DM. Indeed, on the one hand, the stars in the halo either come from the tidal stripping of accreted satellites, or from star formation along gas filaments connected to the halo. In both cases, their kinematics should be imprinted by the kinematics of the gas from which they were formed, which shocks and forms stars with well-ordered motions, near the bottom of the large scale structure's potential wells. On the other hand, the \DM\@ shell-crosses but never shocks, so that particle motions are less ordered, whether the \DM\@ halo comes from the first galactic gravitational collapse, the accretion of satellites or slower accretion from \DM\@ filaments \citep{Pichon+2011, Stewart+2011,Danovich+2015}. We therefore test two additional values of the stellar halo's anisotropy: one tangentially anisotropic ($\beta=-0.8$), and one radially anisotropic ($\beta=0.49$), which lies near the maximum central anisotropy allowed for a Hernquist sphere \citep[see][]{An+2006} and is closer to the stellar halo anisotropy measured by \cite{Bird2019}. In both cases, such an anisotropic distribution may also describe a sub-population of the stellar halo, e.g. stars accreted from a single merger event \citep{Belokurov+2018}, or from a collection of satellites \citep{Riley+2019}. 

\begin{figure}
\centering
    \begin{minipage}{0.48\textwidth}
    \centering
    \includegraphics[width=0.85\textwidth]{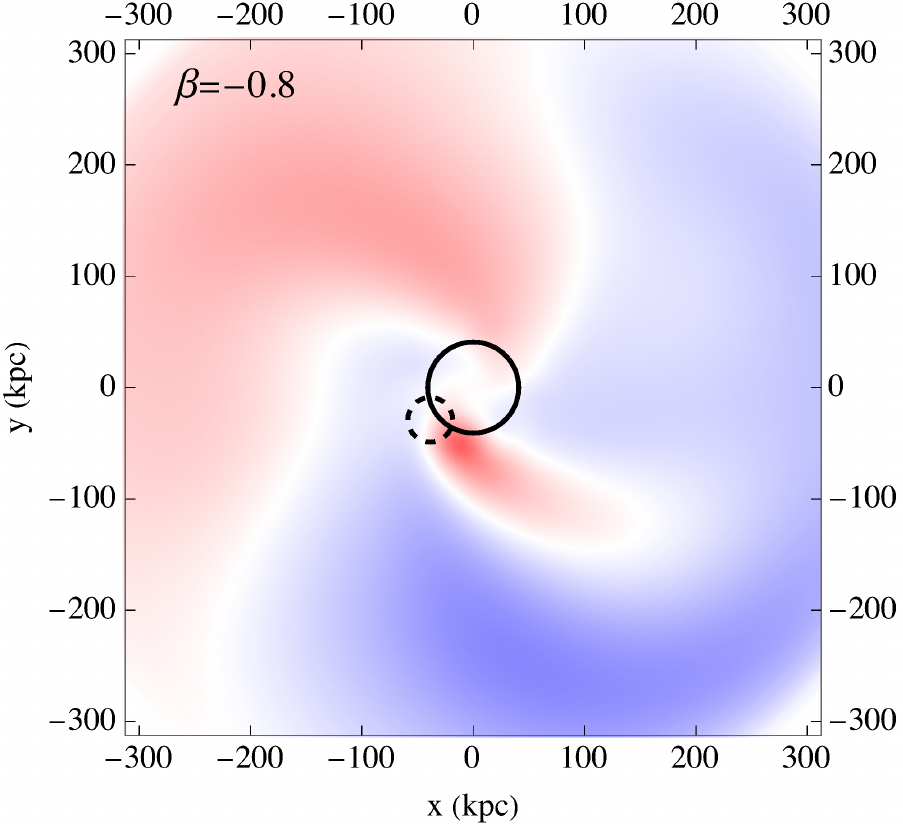} 
    \hspace{-0.2cm}
    \includegraphics[width=0.12\textwidth]{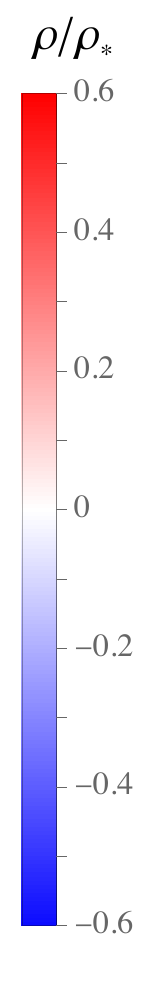}
    \vspace{-0.3cm}
    \end{minipage}
    \begin{minipage}{0.48\textwidth}
    \centering
    \hspace{-0.18cm}
    \includegraphics[width=0.85\textwidth]{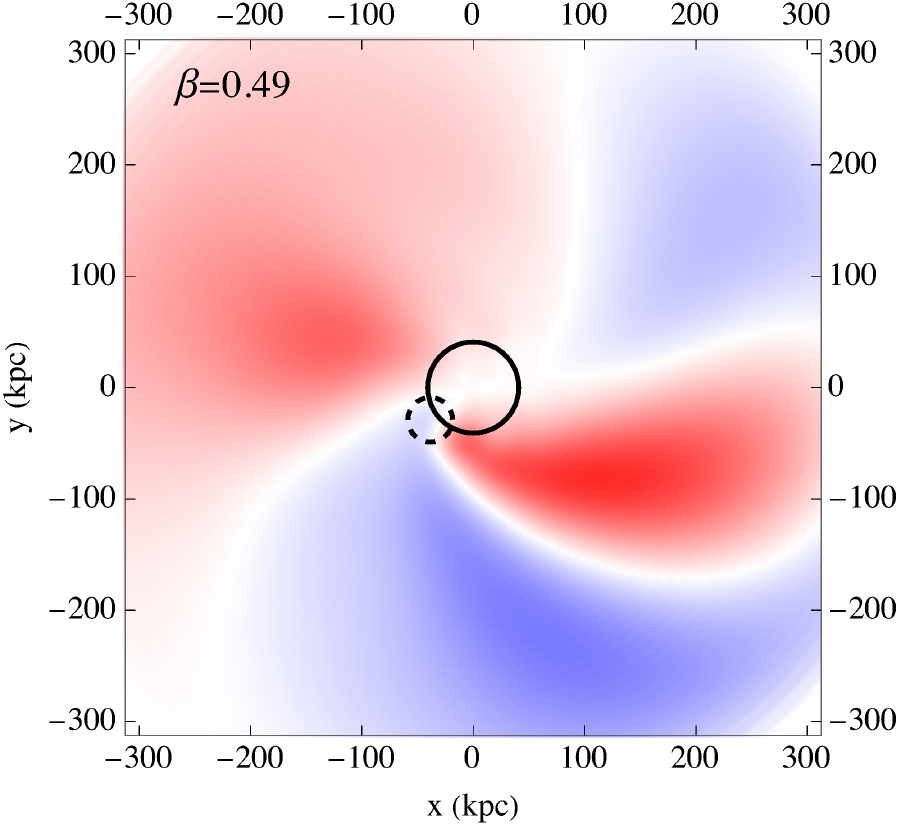} 
    \hspace{-0.2cm}
    \includegraphics[width=0.12\textwidth]{barResponseBeta0.49.pdf}
    \end{minipage}
    \vspace{-0.2cm}
    \caption{Response of the \MW\@ stellar halo, when different anisotropies of the halo are considered: tangential anisotropy ($\beta = -0.8$, top panel) and radial anisotropy ($\beta = 0.49$, bottom panel). The response of the \MW\@ stellar halo is much stronger when it is radially anisotropic. }
\label{fig:MW_Stellar_Halo_Response}
\end{figure}

Figure~\ref{fig:MW_Stellar_Halo_Response} presents the results of this experiment: it shows the relative overdensity of the stellar halo at the present time in the orbital plane of the \LMC\@ for both values of the stellar halo's anisotropy. These panels clearly show that the response is much stronger when the stellar halo is radially anisotropic than when it is tangentially anisotropic. In more detail, it appears that the large dipole is slightly depleted in the radially anisotropic sphere, while the small scale wake is strongly amplified in that same case, and strongly depleted in the tangentially anisotropic system. Finally, in the radially anisotropic system, a small scale overdensity appears in the region of space preceding the \LMC\@ on its orbit, while this region of space presents a small scale underdensity in the tangentially anisotropic stellar halo.


\section{Discussion} 
\label{sec:Discussion}

The results of Section~\ref{sec:MWResponse} showed that the matrix method is able to reproduce $N$-body experiments, and those of Section~\ref{sec:Response_Anisotropic_Halo} clearly show that the stellar halo response depends on its initial anisotropy. Let us now discuss how these results depend on the specific orbit considered for the \LMC, and how they provide physical insight on the \MW's response, as well as quantitative constraints on the stellar halo's anisotropy.

\subsection{Influence of the \LMC's orbit}
\label{subsec:Influence_LMC_Orbit}

We now explore the influence of the specific \LMC\@ orbit we chose for our previous computations. Indeed, the orbit was computed by integrating backwards a point-mass in the \MW\@ static Hernquist potential, which makes it unrealistic in several ways. First, the \LMC\@ is not a point mass, and its orbit is influenced by the fact that the \LMC\@ and the \MW\@ are two inter-penetrating extended objects. More importantly, the \MW\@ is not static, and its intricate reflex motion strongly affects the \LMC\@ dynamics in the frame centered on the \MW\@ cusp.

In order to build a more realistic orbit, we integrated the motion of the \LMC-\MW\@ system using a leap-frog integrator with the following prescriptions:
\begin{itemize}
    \item The motion of the \MW\@ is represented by that of its cusp: at each time step, the \MW's acceleration is that of a point-mass in the \LMC's potential, so that
    \begin{equation}
        \bf a_{\mathrm{MW}} = - \mathbf \nabla \psi_{\mathrm{LMC}}(\mathbf{x}_{\mathrm{MW}}).
        \label{eq:MW_Acceleration_New_Orbit}
    \end{equation}
    In this expression, $\psi_{\mathrm{LMC}}$ represents the current \LMC\@ potential, in particular it depends on its current position.
    \item The motion of the \LMC\@ is computed within the Hernquist \MW\@ potential, but centered on its cusp as determined by the previous step. The acceleration is then
    \begin{equation}
        \mathbf{a}_{\mathrm{LMC}} = - \frac{1}{\MLMC} \int \!\! \rd^3 x \,  \mathbf{\nabla} \psi_{\mathrm{MW}} (\mathbf x) \, \rho_{\mathrm{LMC}} (\mathbf x).
        \label{eq:LMC_Acceleration_New_Orbit}
    \end{equation}
\end{itemize}
Note that these rules differ from those of \cite{Gomez2015}, who also consider the \LMC\@ to be a point-mass in the \MW\@ potential to compute $\mathbf{a}_{\mathrm{LMC}}$. The present prescriptions were chosen because they match more closely the physical processes at play, as well as the framework of the matrix method. Indeed, choosing the \MW\@ cusp as the reference for computing forces is relevant for two reasons. On the one hand, since the satellite is at its first infall, it is always sensitive to more and more central parts of the \MW\@ -- the region of the cusp --, while the \MW's outskirts act as a spherically symmetric shell with no resultant gravitational influence. In the central region, the potential is still that of a Hernquist sphere, centered on the cusp, hence we take that potential as that responsible for the \LMC's acceleration. On the other hand, the matrix method computes the response of the \MW\@ in the reference frame of its cusp, so this refined prescription for the \LMC's orbit is more adapted to it. 

Note that this prescription does not conserve momentum, as the reciprocal forces applied by one object on the other are not equal. However, it appears that it still produces quasi-periodic trajectories. Using the same values of the \LMC\@ pericentric radius and velocity as in Section~\ref{subsec:Models_MW_LMC}, we obtain the orbit shown in Fig.~\ref{fig:LMC_orbit_distance}. This new orbit is consistent with the new prescriptions we applied to the problem: since the \MW\@ is now moving in the \LMC's potential, the \LMC\@ should come from further away in order to produce the same pericentric distance.

\begin{figure}
\centering
    \begin{minipage}{0.48\textwidth}
    \centering
    \includegraphics[width=0.85\textwidth]{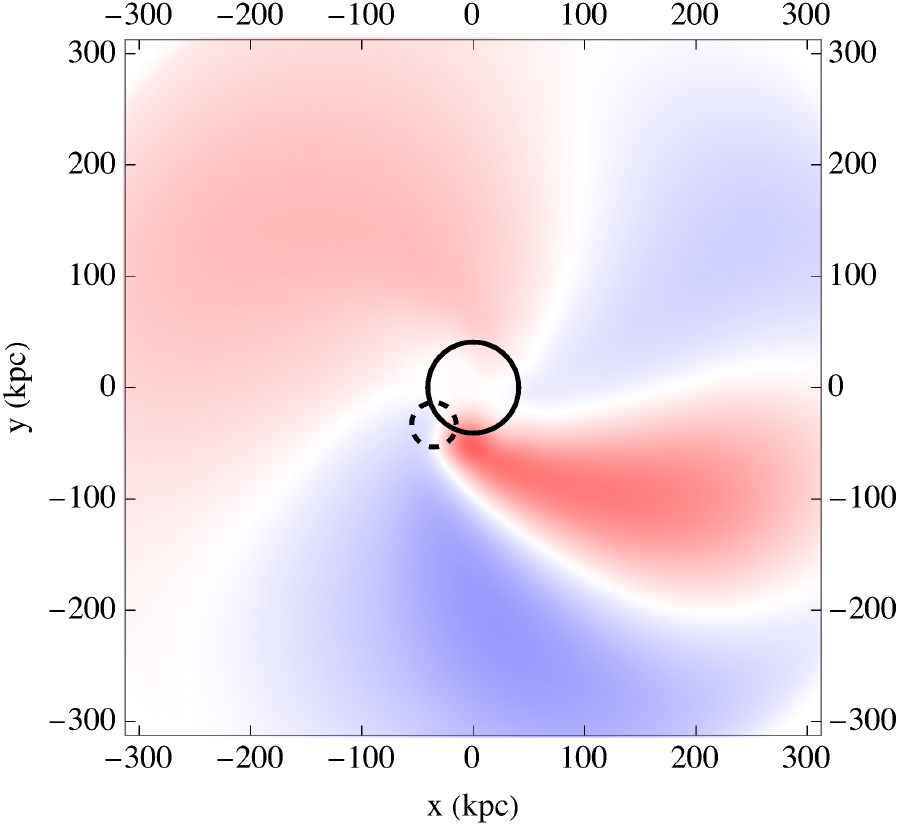} 
    \hspace{-0.2cm}
    \includegraphics[width=0.13\textwidth]{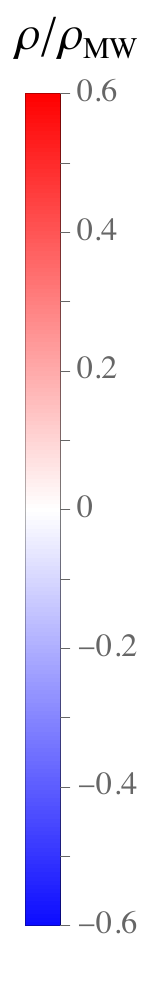}
    \end{minipage}
    \vspace{-0.2cm}
    \caption{Density response, in the orbital plane, at the final time step ($t_{20} = 2 \, \mathrm{Gyr}$) of the fiducial \MW\@ ($\beta = 0$) to the \LMC\@ following the orbit described by eqs.~\eqref{eq:MW_Acceleration_New_Orbit} and~\eqref{eq:LMC_Acceleration_New_Orbit}. The \MW's response is only weakly impacted by the details of the \LMC's orbit at early stages of its infall.}
\label{fig:Response_New_Orbit}
\end{figure}

We then computed the response of the fiducial \MW\@ ($\beta = 0$) to this new orbit for the \LMC. Let us emphasize that this does not require us to re-compute the response matrix, but only to apply eq.~\eqref{eq:Linear_response_final} with the new perturbing vector. In order to compare the present case with the response of Section~\ref{sec:MWResponse}, we consider here the full self-gravitating response of the \DM\@ + stellar halo to the \LMC\@ on its new trajectory. The results are shown in Fig.~\ref{fig:Response_New_Orbit}. Comparing this map with the last panel of Fig.~\ref{fig:MW_response_evolution}, it appears that the \MW's response is weakly sensitive to the details of the \LMC's orbit. Only at large radii, in the tail of the local wake (around $(x,y) = (300 \, \mathrm{kpc}, -100 \, \mathrm{kpc})$) and the orientation of the dipolar component can one notice small deviations in the response's shape. This result hinders the possibility to set strong constraints on the \LMC's orbit far in the past from the sole study of its present influence on the \MW.

\subsection{The reflex motion is pure potential}
\label{subsec:Reflex_Motion}

Let us now decompose the \MW\@ stellar halo's response to the \LMC, and disentangle the physical processes responsible for its characteristic shape. In Appendix~\ref{subsec:Spherical_matrix}, we show that the linear response of a spherical, non-rotating system involves no couplings between different $m$ and $\ell$ orders in the multipolar expansion of the perturber. This is particularly highlighted by the $\delta_{m^p}^{m^q} \, \delta_{\ell^p}^{\ell^q}$ term in eq.~\eqref{eq:Response_matrix_spherical}. The response can therefore be decomposed into a sum of separate harmonic terms. Here, we focus on the dipolar over/underdense pattern, while the local wake is studied in Section~\ref{subsec:Quadrupolar_Response}.

\begin{figure}
\centering
    \begin{minipage}{0.48\textwidth}
    \centering
    \includegraphics[width=0.85\textwidth]{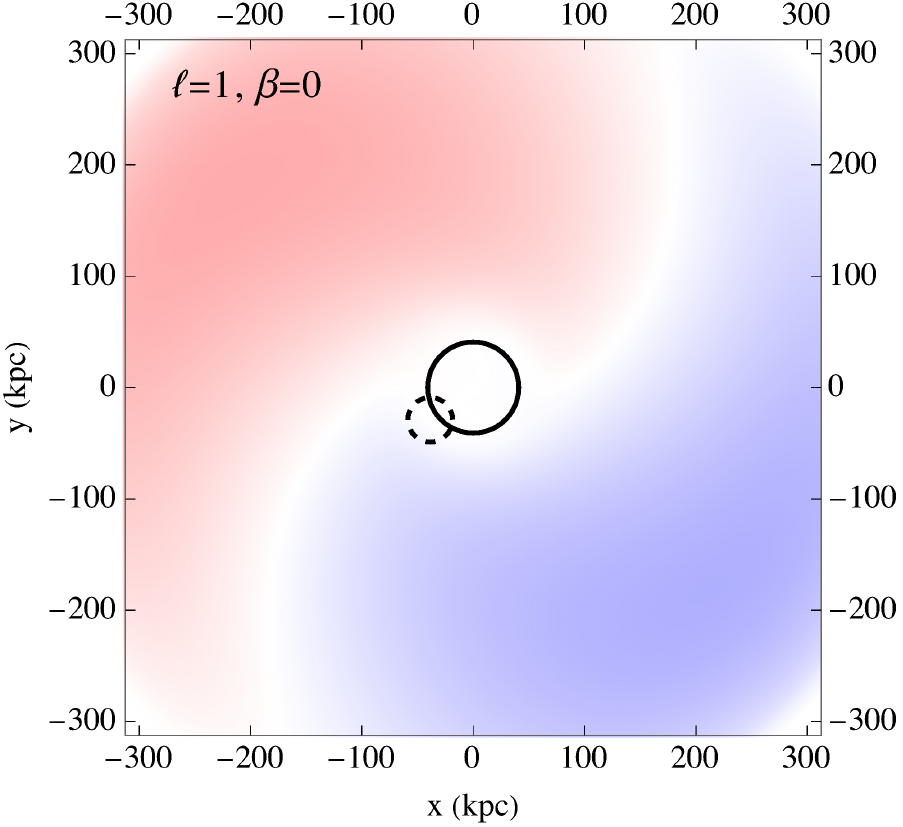} 
    \hspace{-0.2cm}
    \includegraphics[width=0.12\textwidth]{barResponseBeta0.49.pdf}
    \end{minipage}
    \vspace{-0.2cm}
    \caption{Linear response of the \MW\@ stellar halo to the dipolar terms in the perturber. This response includes both $\ell = 1$, $m = \pm 1$ terms. These terms clearly account for the large scale over/underdense pattern related to the differential reflex motion of the \MW\@ halo.}
\label{fig:MW_Dipolar_Response}
\end{figure}

\begin{figure*}
\centering
    \begin{minipage}{0.97\textwidth}
    \centering
    \includegraphics[width=0.93\textwidth]{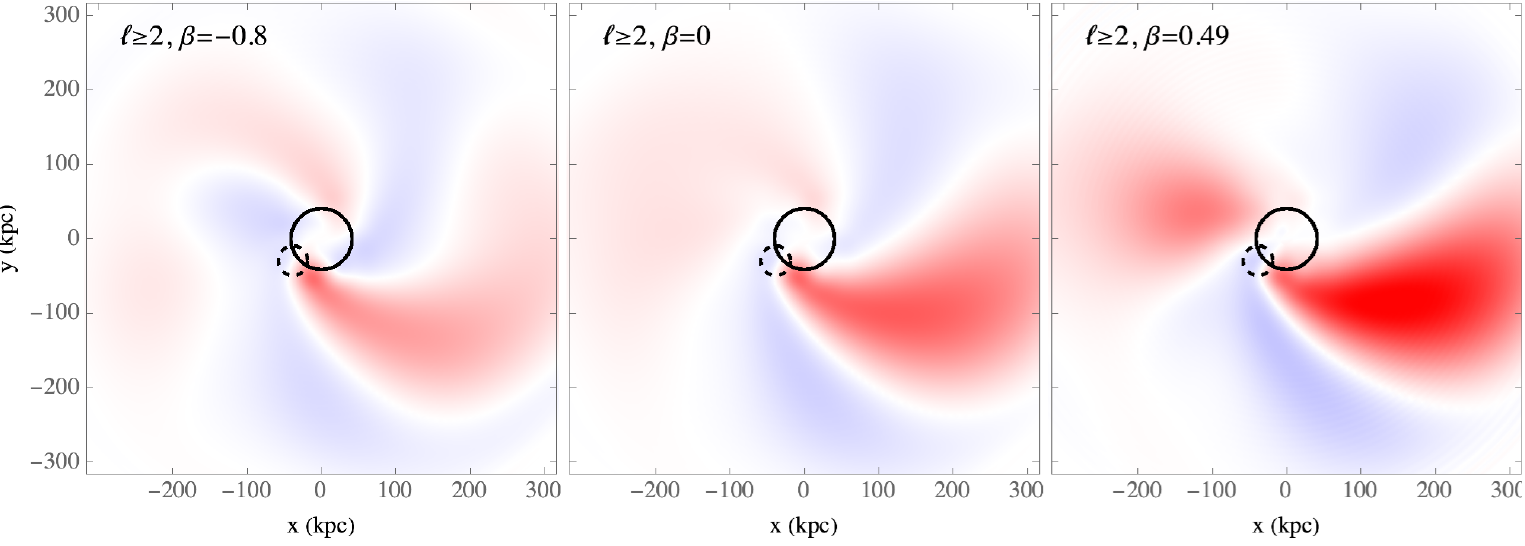} 
    \includegraphics[width=0.052\textwidth]{barReflexResponseRadial.pdf}
    \end{minipage}
    \caption{Linear response of the different \MW\@ models (tangentially anisotropic, isotropic and radially anisotropic, respectively in the left, central and right panels) to the $\ell \geq 2$ terms in the perturber. These terms clearly account for the overdense wake trailing behind the \LMC, as well as other response components unrelated to the \MW\@ reflex motion.}
\label{fig:MW_Wake_Response}
\end{figure*}

Figure~\ref{fig:MW_Dipolar_Response} shows the bare response of the \MW\@ halo to the $\ell = 1$ components of the \LMC. It appears that this harmonic fully accounts for the effect of the differential reflex motion of the \MW: it presents the same dipolar feature, with similar amplitude and rotation. As will be further shown in Fig.~\ref{fig:MW_Wake_Response}, the other harmonics do not contribute to that component in the response, but to the overdense wake trailing behind the \LMC's trajectory.

\begin{figure}
\centering
    \begin{minipage}{0.48\textwidth}
    \centering
    \includegraphics[width=0.85\textwidth]{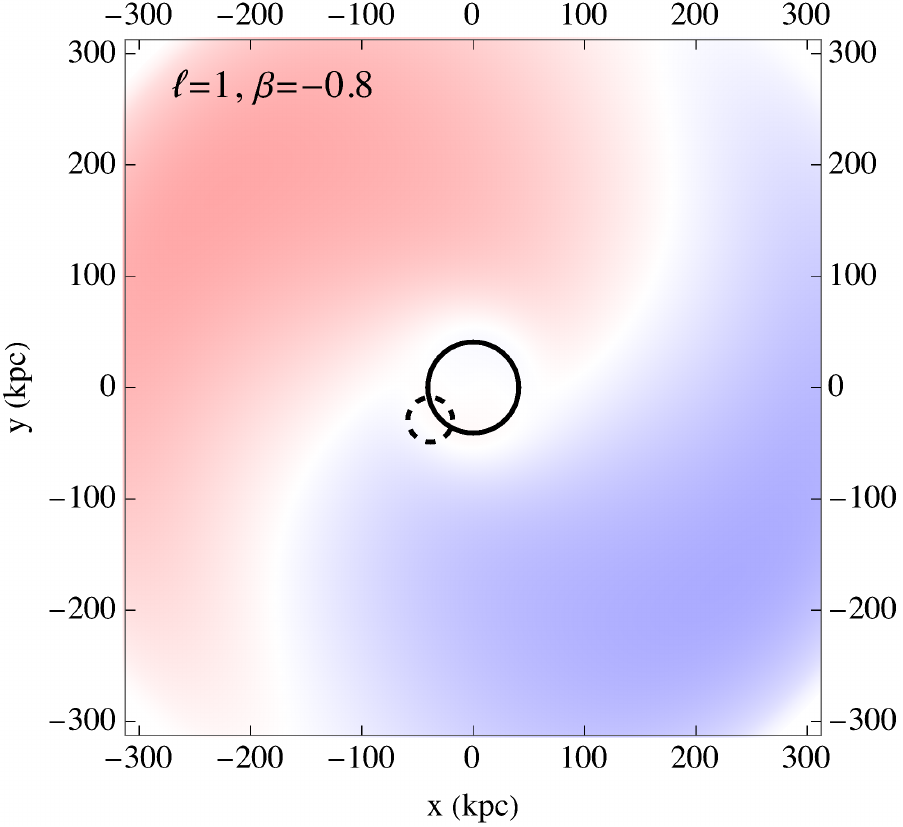} 
    \hspace{-0.2cm}
    \includegraphics[width=0.12\textwidth]{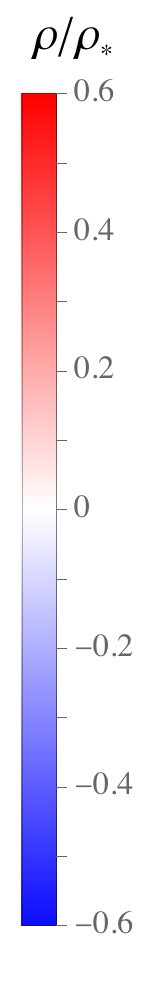}
    \vspace{-0.3cm}
    \end{minipage}
    \begin{minipage}{0.48\textwidth}
    \centering
    \hspace{-0.18cm}
    \includegraphics[width=0.85\textwidth]{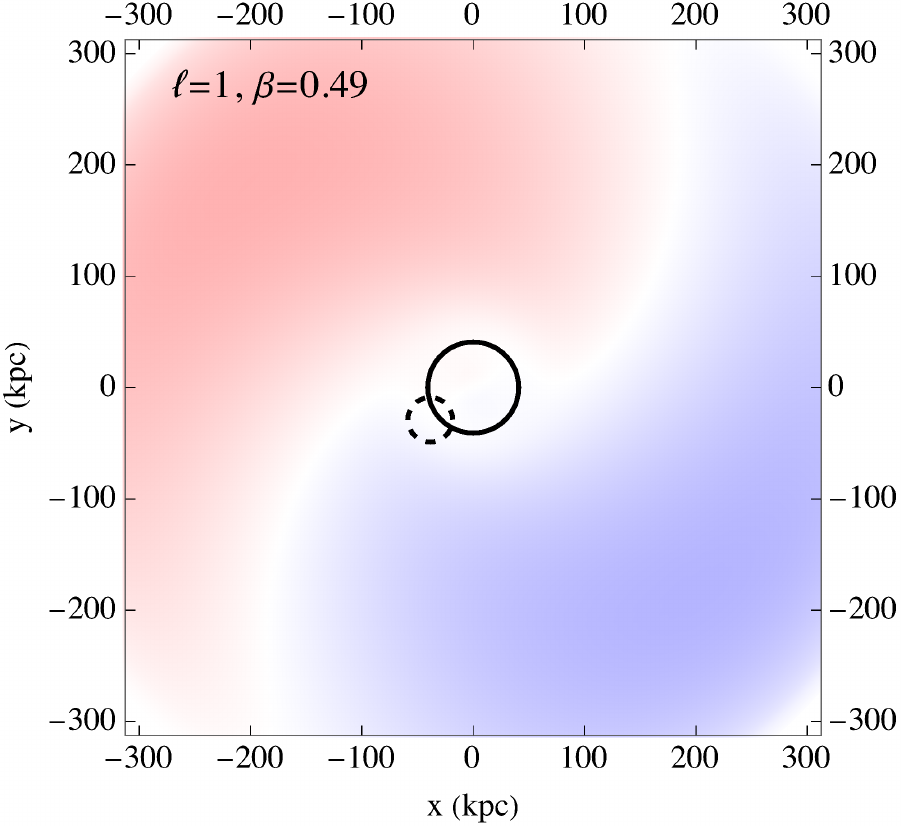} 
    \hspace{-0.2cm}
    \includegraphics[width=0.12\textwidth]{barReflexResponseRadial.pdf}
    \end{minipage}
    \vspace{-0.2cm}
    \caption{Response of the \MW\@ stellar halo to the $\ell = 1$ components of the \LMC, when different anisotropies of the halo are considered: tangential anisotropy ($\beta = -0.8$, top panel) and radial anisotropy ($\beta = 0.49$, bottom panel). The dipolar response of the \MW\@ stellar halo is essentially insensitive to the anisotropy.}
\label{fig:MW_Stellar_Halo_Dipolar_Response}
\end{figure}

Now, let us evaluate the bare response of stellar halos with strong anisotropy to the dipolar components of the \LMC. The results for the radially and tangentially anisotropic halos are shown in Fig.~\ref{fig:MW_Stellar_Halo_Dipolar_Response}. One can clearly see the similarity between the present maps and Fig.~\ref{fig:MW_Dipolar_Response}: both the amplitude of the dipolar response and its shape do not seem to depend on the velocity anisotropy of the \MW\@ stellar halo. This implies that the \MW\@ reflex motion is independent of its internal kinematics, and only depends on its potential and on the characteristics of the \LMC's orbit. 

Note that our ability to study this dipolar term using the matrix method was provided by a simple yet critical improvement, i.e.\@ by considering the motion in the reference frame of the density maximum instead of the \COM. Indeed, previous studies \citep[see related discussions in][]{Seguin1994,Murali1999} were considering the \COM\@ as the reference frame for matrix computations, so that the density maximum gets displaced from the centre of the reference frame. This has two unfortunate consequences on the ability of the matrix method to accurately reconstruct dipolar effects. On the one hand, the central displacement appeared as a rather large amplitude dipolar pattern located close to the centre. This made it difficult to reconstruct it using bi-orthogonal basis functions, all the more that there is a large contrast between the central density and its close vicinity (e.g., in a cuspy system). This central displacement could also dominate over the more subtle effects related to the host's reflex motion, which we study in more detail here. On the other hand, the matrix method makes the assumption that the potential centre is at the reference frame centre, any deviation being considered as a small perturbation. When the density maximum is displaced, the linear hypothesis may be violated, all the more so for a cuspy system. All in all, working in a reference frame which follows the density maximum at each time addresses some significant problems.
The problem of the density maximum being off-centered is also a concern to $N$-body codes using basis function expansions (e.g., the so-called self-consistent field codes). This problem is usually dealt with by re-centering the origin of the expansion at the density maximum of the particles \citep[see, e.g.,][]{Choi2007,Meiron2014}, but these codes may also benefit from the method we developed here to follow more closely the density maximum and correct the force calculations accordingly.

\subsection{The local wake as a probe of the halo anisotropy}
\label{subsec:Quadrupolar_Response}

Let us now focus on the $\ell \geq 2$ part of the response. The central panel of Fig.~\ref{fig:MW_Wake_Response} shows the response of the isotropic \MW\@ to the $\ell \geq 2$ components of the \LMC. Clearly, this part of the response encompasses the local overdensity trailing behind the \LMC. One can also see traces of a slight overdensity on the other side of the \MW, as well as underdense regions between these two overdensities. However, these last two features are much less significant in these regions than the dipolar pattern related to the reflex motion. 

If we consider Fig.~\ref{fig:MW_Wake_Response} altogether, we can evaluate the influence of the stellar halo anisotropy on the aspect of the local wake. Clearly, its amplitude is increased in the radially anisotropic halo, and depleted in the tangentially anisotropic one. The wake's structure is also different: in the tangential halo, the wake follows very closely the \LMC\@ on its orbit, while there is an underdensity ahead of the satellite. In the radial case, however, the part of the wake trailing behind the \LMC\@ does so from a distance, and is accompanied by an important overdensity directly ahead of the satellite on its orbit.

\begin{figure*}
\centering
    \begin{minipage}{0.97\textwidth}
    \centering
    \includegraphics[width=0.93\textwidth]{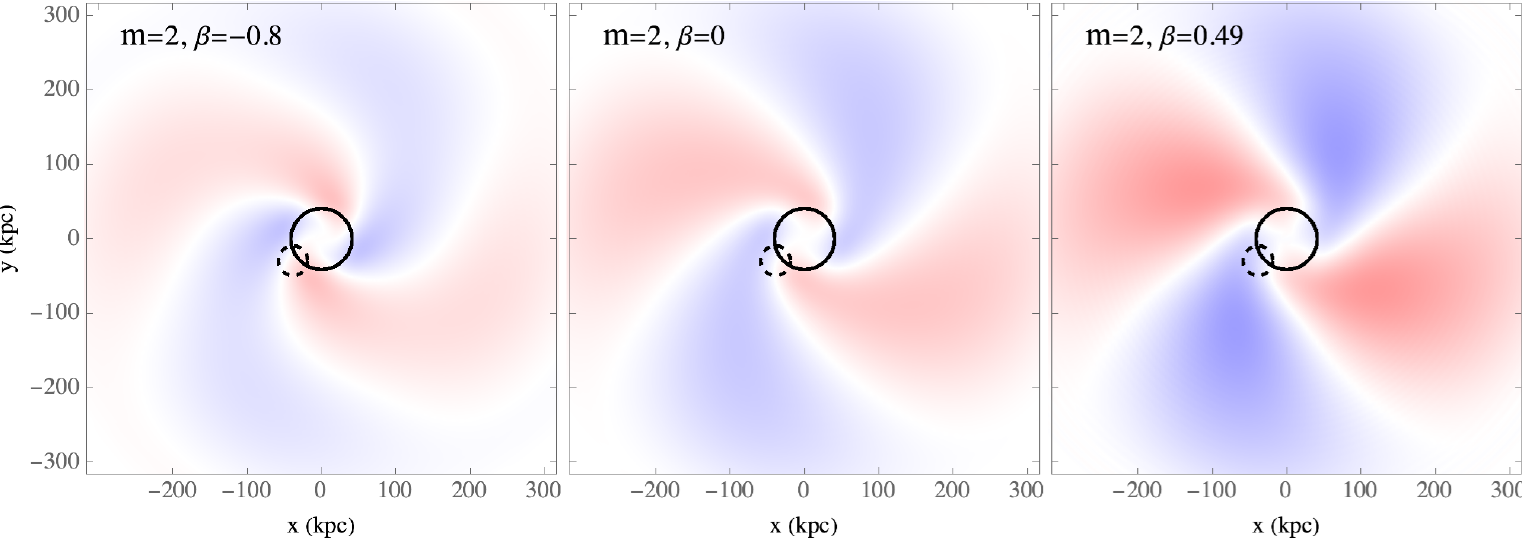}
    \includegraphics[width=0.052\textwidth]{barReflexResponseRadial.pdf}
    \end{minipage}
    \caption{Linear response of the different \MW\@ models (tangentially anisotropic, isotropic and radially anisotropic, respectively in the left, central and right panels) to the $m=2$ terms in the perturber. The orientation of this quadrupolar response can represent a crucial probe of the stellar halo's anisotropy.}
\label{fig:MW_Quadrupolar_Response}
\end{figure*}

As a consequence, it appears that the anisotropy of the stellar halo can be probed by measuring the structure and amplitude of the wake in the stellar density of the halo. Let us be more specific, and focus on a possibly measurable feature: the quadrupolar ($m=\pm2$) component of the stellar distribution in the orbital plane of the \LMC. Indeed, this quantity could be recovered as a Fourier transform of the stellar halo density along the angular direction in that plane. Notice that, in our response, these terms involve all even $\ell \geq 2$ harmonics with $m=\pm2$. Figure~\ref{fig:MW_Quadrupolar_Response} shows the quadrupolar response of stellar halos with different anisotropies. Not only does the amplitude of this term vary with anisotropy, but more importantly, it is more wound in the tangential halo than in the radial one, which entails different orientations in the central parts. Within $\sim 50 \, \mathrm{kpc}$, the positive part of the $m=\pm2$ response is nearly aligned with the position of the \LMC\@ in the case of a tangential halo, while it is largely misaligned from that position if the halo is radial. Further away, the orientation of this harmonic evolves at a greater pace in the tangential case, so that in all cases, its orientation at large distances coincides with the initial location of the \LMC.

The exact angles between this quadrupolar response and the \LMC\@ are likely to also depend on the \MW\@ potential and on the details of the \LMC's orbit. However, there is little doubt that this dependence of the orientation on anisotropy will remain in other models of the \MW-\LMC\@ interaction. Furthermore, we also expect that other kinematic features of the stellar halo (e.g., rotation, or different distributions of the anisotropy) would imprint its quadrupolar response.

\subsection{Frequencies in the response}
\label{subsec:Resonant_Effects}

\begin{figure}
\centering
    \begin{minipage}{0.48\textwidth}
    \centering
    \includegraphics[width=0.85\textwidth]{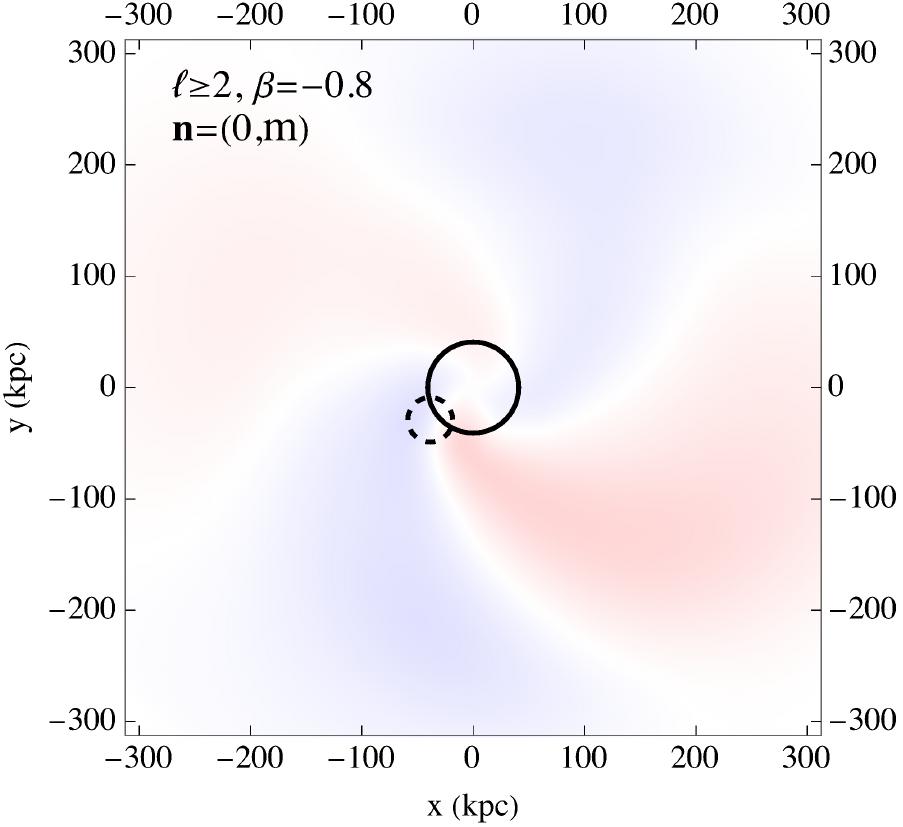} 
    \hspace{-0.2cm}
    \includegraphics[width=0.12\textwidth]{barReflexResponseRadial.pdf}
    \vspace{-0.3cm}
    \end{minipage}
    \begin{minipage}{0.48\textwidth}
    \centering
    \hspace{-0.18cm}
    \includegraphics[width=0.85\textwidth]{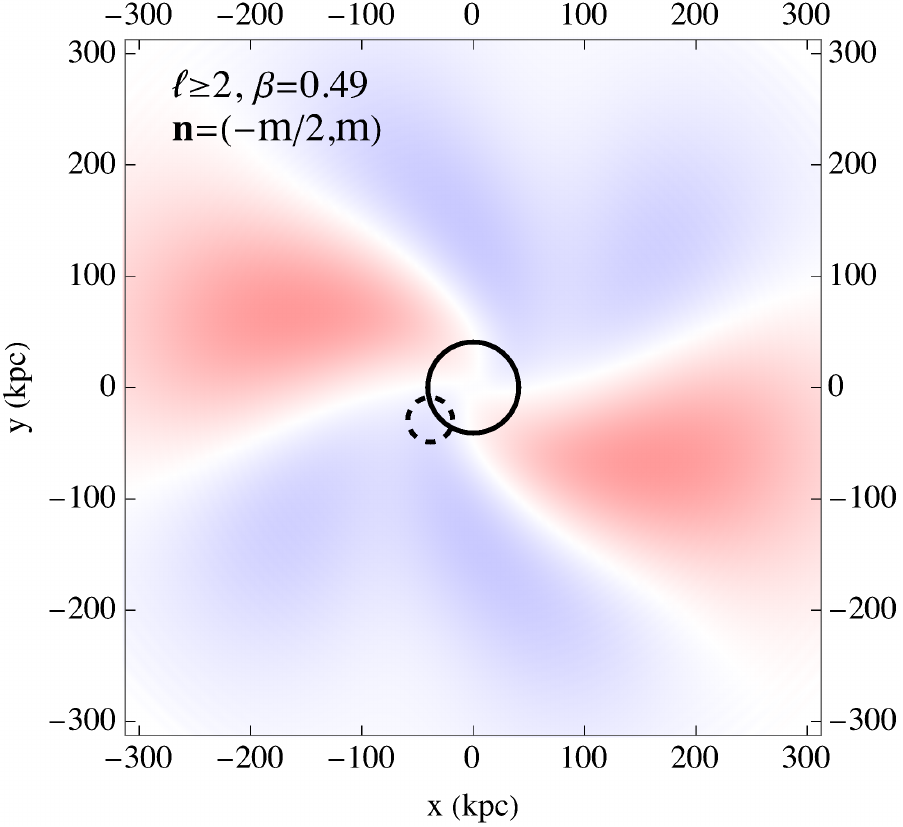} 
    \hspace{-0.2cm}
    \includegraphics[width=0.12\textwidth]{barReflexResponseRadial.pdf}
    \end{minipage}
    \vspace{-0.2cm}
    \caption{Response in the dominant resonant contribution in the tangential (top) and radial (bottom) models of the \MW\@ stellar halo. In the tangential case, the dominant resonance is the corotation, characterised by $\bn = (0,m)$. The response's shape is similar to the full wake, but its amplitude is much lower: many other combinations of frequency have an important contribution to the response. In the radial case, the dominant resonance is the inner Lindblad resonance, characterised by $\bn = (-m/2, m)$. The response's shape is similar to the full wake, and its amplitude is about 50\% of the wake: even though other frequencies also contribute, this resonance is clearly dominant.}
\label{fig:Resonant_Response}
\end{figure}

A possible advantage of the matrix method, compared to $N$-body simulations, is to interpret the \MW's response as the combined action of orbital frequencies. Indeed, since the bare response of the stellar halo merely involves the application of the response matrix to the external perturber (see eq.~\eqref{eq:Bare_Response}), it can be straightforwardly decomposed into the sum of the influence of different combinations of orbital frequencies, corresponding to each value of the Fourier numbers $\bn$ (also called resonance vector) in eq.~\eqref{eq:Response_Matrix}\footnote{This analysis is coined the \textit{restricted matrix method} in \cite{Breen2021}. In the present case, the absence of self-gravity makes the interpretations easier, because the resonant terms are directly additive.}. As we will see, comparing their influence yields very distinct pictures in the radially and tangentially anisotropic models of the \MW. 

The influence of each $\bn$ component on the perturber $\bb$ can be interpreted as follows. Let us decompose the perturber spatially in each of its azimuthal harmonics $m$, then further decompose one of these spatial harmonic components through a time Fourier transform with a spectral continuum of frequencies $\op$. The resulting pattern appears as an $m$-armed feature in the $z=0$ plane, with pattern speed $\Op = \op / m$ \citep[see, e.g.,][]{TremaineWeinberg1984}. A single of these components therefore follows
\begin{equation}
    \bb_m(t) = \bb_{m0} \, \re^{- \ri m \Op t}.
\end{equation}
When a single $\bn$ component of the response matrix is applied to this periodic perturber, it yields the bare response (see eqs.~\eqref{eq:Response_time_integral} and~\eqref{eq:Response_Matrix})
\begin{equation}
\ba_m(t) \!=\! \!\!\int_0^t \!\!\!\! \rd \tau \!\!\int\!\! \rd \bJ \, \re^{- \ri \, \bn \cdot \bO \, t} \, \re^{\ri \, (\bn \cdot \bO  - m \Op)\, \tau} \, \bN_{\bn}(\bJ) \bb_{m0},
\label{eq:Response_Single_Frequency}
\end{equation}
where the matrix $\bN_{\bn}(\bJ)$ is defined in terms of its components by
\begin{equation}
    \bN_{\bn, pq}(\bJ) \!=\! - \ri (2 \pi)^3 \bn \!\cdot\! \frac{\partial F}{\partial \bJ}  \psi_{\bn}^{(p)*\!}(\bJ)  \big(\psi_{\bn}^{(q)\!}(\bJ) \!+\!  \phi_{\bn}^{(q)\!}(\bJ) \big).
\end{equation}
Now, the integral over $\tau$ in eq.~\eqref{eq:Response_Single_Frequency} can be straightforwardly performed, giving
\begin{align}
\nonumber
    \ba_m(t) = t \!\!\int\!\! \rd \bJ \, \re^{- \ri \tfrac{\bn \cdot \bO + m \Op}{2} t} \, & \sinc\big(\tfrac{\bn \cdot \bO - m \Op}{2} t\big) \\
    & \times \bN_{\bn}(\bJ) \bb_{m0}.
    \label{eq:Response_Toy_Model_Resonances}
\end{align}
This expression helps us identify the orbits which will have a large impact on the response in terms of their orbital frequencies. Indeed, when the phase space is scanned by the integral over $\bJ$, the $\sinc$ function acts as a window which selects the orbits for which ${\bn \cdot \bO - m \Op \simeq 0}$, all the more that time gets large. This implies that the component we selected in the perturber (with space-time structure given by $m$ and $\op$) has a larger impact on regions of phase space where $\bn \cdot \bO - m \Op \simeq 0$ for one of the Fourier sets $\bn$. This motivates calling the Fourier numbers $\bn$ a resonance vector, and considering the resonance condition 
\begin{equation}
\bn \cdot \bO = m \Op  
\label{eq:Resonance_Condition}
\end{equation}
in the following discussion. Additionally, this model gives us insight into how the duration of the merger impacts the response. At early times, the frequency spectrum of the perturber is rather broad, and the width of the sinc window is too, so that many different regions of the host's frequency distribution can interact with the perturber. At later times, the perturber's frequency spectrum narrows down to perhaps a single or a set of frequencies (because it follows a quasi-periodic orbit). Besides, the width of the sinc window gets narrower too. Therefore, we can expect resonances to play a more prominent role on longer time scales.

In the tangential case, it appears that several combinations of frequencies add up to produce the final wake, the most important of which being the corotation frequency. The top panel of Fig.~\ref{fig:Resonant_Response} shows the final response of the \MW\@ when only $(n_1, n_2) = (0, m)$ terms are kept in the response, for all $|m| \geq 2$. 
We associate those $(n_1, n_2) = (0, m)$ combinations with corotating orbits, corresponding to stars which rotate together with the perturber, at the same orbital rate $\Omega_2 = \Op$ (see eq.~\eqref{eq:Resonance_Condition}). In Fig.~\ref{fig:Resonant_Response}, it appears that the contribution of this frequency accounts for the bulk of the wake, with the same shape, the same position but a lower amplitude. On top of it, a number of smaller contributions from other frequencies add up to form the full wake of Fig.~\ref{fig:MW_Wake_Response}. It should be noted that the corotation frequency has a relatively high value, which could be related to the fact that (i) the wake is able to move fast and to follow the perturber closely, and (ii) the response dissipates efficiently once the \LMC\@ enters the fastest portion of its orbit, which produces a rather shallow wake at the last time step.

In the radial halo, the wake is clearly dominated by the contribution of the inner Lindblad frequency, $\OILR = \Omega_2 - \Omega_1 / 2$. Indeed, the bottom panel of Fig.~\ref{fig:Resonant_Response} shows the response when $(n_1, n_2) = (-m/2, m)$ for $|m| = 2, 4$, which accounts for most of the amplitude of the full wake of Fig.~\ref{fig:MW_Wake_Response}. These combinations of frequencies were selected because they correspond to $\OILR = \Op$ in eq.~\eqref{eq:Resonance_Condition}. The fact that this specific combination of frequencies dominates tends to indicate that instead of attracting particles which can move with it, the \LMC\@ rather attracts orbits which can precess with it. This $\OILR$ frequency has a smaller value than the orbital frequency $\Omega_2$, which could explain the fact that (i) the wake is relatively slow and trails behind the \LMC\@ from a distance, and (ii) the overdensity appearing at early times, when the \LMC\@ is slow and able to attract lots of material, never dissipates, so that the final wake has a large contribution from this epoch.

The influence of resonant frequencies should however be interpreted with care, for several reasons. First, the building blocks of the matrix method are orbits and frequencies, which may over-emphasise the role of resonances in every linear problem. For example, dominant resonant effects can also be identified when the same analysis is applied to the reflex motion of Section~\ref{subsec:Reflex_Motion}, although it does not appear as a resonant process. Second, the resonant interpretation is based on the assumption that the frequency spectrum of the perturber is narrow, which is obviously not the case because the \LMC\@ is integrated for just a fraction of its orbit. 
Finally, the \MW's response is only integrated for $2 \, \text{Gyr}$, which is quite short compared to the orbital times in the stellar halo, whereas the effect of resonances could only truly dominate the \MW’s response over secular timescales. On shorter timescales, their impact is broadened by a width given the interaction’s timescale (see eq.~\eqref{eq:Response_Toy_Model_Resonances}). 

To summarize these arguments, we propose the following interpretation, also consistent with the results of \cite{Seguin1994,Weinberg1989}. In the very early stages of the interaction ($t \lesssim 1 \, \text{Gyr}$), the stars in the \MW\@ halo behave like pure particles and are attracted by the gravitational field of the slowly moving \LMC. In an idealised future, the \LMC\@ will have made several rotations around the \MW\@ on its orbit, and the commensurabilities between the orbits of stars in the \MW\@ and the \LMC's orbit could be the main driver of the \MW\@ stellar halo's response. In the meantime, the situation is mixed: the initial impulsive effects are still important, but the stars also started to feel the frequency structure of the \LMC's motion.

\section{Conclusion} 
\label{sec:Conclusion}

Recent photometric and spectroscopic surveys focusing on the \MW's stellar halo showed that it is dynamically perturbed by the influence of the \LMC. These observations require detailed modelling of this infall, in order to extract constraints on its characteristics: density and kinematics of the \MW's dark matter and stellar halos, mass and orbit of the \LMC. While this problem is mostly studied by means of $N$-body simulations, we took here a complementary approach relying on the matrix method from linear response theory, which yields an explicit estimator for the impact of the \LMC\@ on the structure of the stellar and \DM\@ halos. We first showed that this analytical tool is able to quantitatively reproduce the results of $N$-body simulations, opening the possibility of framing the match to the observed response of the stellar halo as an inverse problem for its internal kinematics. In addition, the matrix method gives interesting insight on the physical processes at play, which informs us on the effect of minor mergers on galaxies in general. Our main results are the following:
\begin{itemize}
    \item[(i)] At that stage of the interaction, the response's self-gravity has little influence. As a consequence, the stellar halo can be considered as a massless component, and its response is weakly sensitive to the \DM\@ halo kinematics.
    \item[(ii)] The response of the \MW\@ stellar halo is strongly dependent on its internal kinematics. Here, we focused on its anisotropy, but it is expected that global angular momentum also would impact its response.
    \item[(iii)] The \MW's response is weakly dependent on the details of the \LMC's orbit. We compared the response when the \LMC's orbit is integrated in a static \MW\@ potential, with that when the \MW\@ centre is also allowed to move in the \LMC's potential, yielding very similar results in the density response of the \MW\@ halo.
    \item[(iv)] The reflex motion of the \MW\@ corresponds to the dipolar $\ell = 1$ part of the response. Its structure is very weakly sensitive to the internal kinematics of the \MW\@ halo, and mainly depends on its potential.
    \item[(v)] The local wake corresponds to the $\ell \geq 2$ part of the response. Its structure is strongly dependent on the stellar halo's anisotropy, both in amplitude and in spatial distribution. Typically, a tangentially anisotropic halo produces a shallow wake and an underdensity ahead of the \LMC\@ on its orbit, while a radially anisotropic halo produces a strong wake and an overdensity ahead of the \LMC.
    \item[(vi)] The winding and orientation of the quadrupolar in-plane response ($m=2$) of the halo represents a novel probe of its anisotropy. Typically, a tangentially anisotropic halo produces a tightlier wound $m=2$ response where the overdensity is nearly aligned with the \LMC\@ within $50 \, \text{kpc}$ of the \MW, while it is more loosely wound and rather misaligned with the \LMC\@ in the radial halo.
    \item[(vii)] Resonances seem to matter, in particular in the radially anisotropic case with radial orbits at the inner Lindblad resonance with the \LMC. 
\end{itemize}

These first results call for further work relying on the linear response formalism. On the one hand, the structure and kinematics of the \MW\@ could be modified, in order to account for the details of the \DM\@ halo density profile (e.g., different central and outer slopes, triaxiality, or clumpiness of the halo), and of the complexity of the stellar halo kinematics (e.g., global angular momentum, spatial distribution of the anisotropy). In particular, it is possible that even a mild level of rotation could significantly impact the \MW\@ halo's response. On the other hand, the tidal evolution of the \LMC\@ could be included in the model, in order to evaluate how the evolution of its mass influences the \MW's response. Surely, such studies will help refining our models of the LMC infall.

The method developed here should also be adapted to disk-halo configurations, in order to allow the detailed analytical study of the response of the \MW\@ disk-halo system to the accretion of the Sagittarius dwarf galaxy. In this merger, the effect of the self-gravity of the \DM\@ halo could potentially play a much more important role because of the multiple wraps that the Sagittarius dwarf has already made around the \MW\@ halo. More generally, it would be important to investigate how and when self-gravity becomes important, depending on the kinematic structure of the host galaxy. 


%

\section*{Acknowledgements}
This work has been supported by the Agence Nationale de la Recherche (ANR project GaDaMa ANR-18-CE31-0006). It has also received funding from the project ANR-19-CE31-0017 and from the European Research Council (ERC grant agreement No. 834148). We used the High Performance Computing facility of the Observatoire astronomique de Strasbourg for our computations, and we thank Mathieu Misslin for running it smoothly.

\bibliography{biblio}{}
\bibliographystyle{aasjournal}



\numberwithin{equation}{section} 
\numberwithin{figure}{section}
\appendix

\section{Derivation of the matrix method}
\label{app:Matrix_method}

Starting from the linearised \CBE\@ (eq.~\ref{eq:CBE_linearised}), let us follow a similar path as \cite{Murali1999,Pichon2006} to derive the response matrix formalism in its temporal version.

\subsection{Angular Fourier transform}

First, let us expand all perturbed quantities of eq.~\eqref{eq:CBE_linearised} in Fourier series of the angles. Since each quantity should be $2\pi$-periodic in the angles, these series can be written as
\begin{subequations}
\label{eq:Fourier_transform_angles}
\begin{equation}
f (\bT , \bJ, t) =  \sum_{\bn} f_{\bn} (\bJ, t) \, \re^{\ri \bn \cdot \bT}, 
\end{equation}
\begin{equation}
f_{\bn} (\bJ, t) = \!\! \int \!\! \frac{\rd \bT}{(2 \pi)^{3}} \, f (\bJ, \bT, t) \, \re^{- \ri \bn \cdot \bT},
\end{equation}
\end{subequations}
and similarly for the perturbing Hamiltonian 
\begin{equation}
\Delta H(\bT , \bJ, t) =  \sum_{\bn} \Delta H_{\bn} (\bJ, t) \, \re^{\ri \bn \cdot \bT}.
\end{equation}
Here, $\bn \in \mathbb{Z}^{3}$ is the triplet labelling each Fourier coefficient. Multiplying eq.~\eqref{eq:CBE_linearised} by $\re^{- \ri \, \bn \cdot \bT}$ and integrating over the angles, each Fourier component separately satisfies
\begin{equation}
\label{eq:CBE_Fourier_transform}
\frac{\partial f_{\bn}}{\partial t} + \ri \, \bn \!\cdot\! \bO \, f_{\bn} = \ri \, \bn \!\cdot\! \frac{\partial F}{\partial \bJ} \, \Delta H_{\bn}.
\end{equation}
Now, eq.~\eqref{eq:CBE_Fourier_transform} takes the form of an integro-differential equation on $f_{\bn}$. Assuming that the system is unperturbed at the initial time, i.e.\ $f_{\bn}(t=0) = 0$, the solution satisfies the integral equation
\begin{equation}
f_{\bn}(\bJ, t) = \ri \, \bn \!\cdot\! \frac{\partial F}{\partial \bJ} \! \int_0^t \!\!\! \rd \tau \, \Delta H_{\bn}(\bJ, \tau) \, \re^{-\ri \, \bn \cdot \bO \, (t - \tau)},
\label{eq:f_Fourier_time_integral}
\end{equation}
where $\Delta H_{\bn}$ itself depends on $f_{\bn}$ through the Poisson equation.

\subsection{Basis function expansion}

To make this dependence explicit, we project the perturbing quantities onto a bi-orthogonal basis of potentials and densities. We can assume that this basis takes the form 
\begin{subequations}
\label{eq:Basis_general_form}
\begin{equation}
\psi^{(p)} (\bx) = \psi_{\ell m n} (r , \theta , \phi) = Y_{\ell}^{m} (\theta , \phi) \, U_{n}^{\ell} (r) ,
\end{equation}
\vspace{-20pt}
\begin{equation}
\rho^{(p)} (\bx) = \rho_{\ell m n} (r , \theta , \phi) = Y_{\ell}^{m} (\theta , \phi) \, D_{n}^{\ell} (r) ,
\end{equation}
\end{subequations}
where a given basis element is characterised by three indices, ${ \ell \geq 0 }$, ${ |m| \leq \ell }$, and ${ n \geq 0}$, and $Y_{\ell}^{m}$ is a spherical harmonic following the normalisation convention
\begin{equation}
\int \! \rd \theta \, \rd \phi \, \sin ( \theta) \, |Y_{\ell}^{m}(\theta, \phi)|^{2} = 1 .
\end{equation}
In eq.~\eqref{eq:Basis_general_form}, we also introduced the radial parts of the respective bases, $U_{n}^{\ell}$ and $D_{n}^{\ell}$, which are normalised so that (see also eq.~\ref{eq:biorthcond})
\begin{equation}
    \!\! \int \!\! \rd \bx \, \psi^{(p)} (\bx) \, \rho^{(q)*} (\bx) = - \delta_p^q.
\end{equation}
Several choices for these radial functions are given in the literature, e.g.\ by \cite{CluttonBrock1973,FridmanPolyachenko1984(II),HernquistOstriker1992,Bertin1994,Zhao1996,RahmatiJalali2009,Lilley+2018}. Unlike what is required to fully reconstruct the perturber and the host \citep[see, e.g.,][]{Weinberg1999}, the matrix method only demands an accurate reconstruction of the perturber, so that there is no need in the present study for a basis with cuspy elements. In Appendix~\ref{app:BOB}, we briefly describe our choice of basis, which is that of \cite{CluttonBrock1973} and has a Plummer profile \citep{Plummer1911} as the first element. 

We define the projections of the perturbing potentials $a_p$ and $b_p$ so that (see also eq.~\ref{eq:Projection_perturbing_potentials1})
\begin{subequations}
\label{eq:Projection_perturbing_potentials}
\begin{equation}
    \psi^{\rs}(\bx, t) = \sum_{p} a_p(t) \, \psi^{(p)}(\bx)
\end{equation}
\vspace{-10pt}
\begin{equation}
    \psi^{\re}(\bx, t) = \sum_{p} b_p(t) \, \psi^{(p)}(\bx)
\end{equation}
\end{subequations}
Using such an expansion, it becomes clear that the purpose of the matrix method will be to compute the coefficients $a_p(t)$ (the system's response), given a certain set of coefficients $b_p(t)$ (the external perturber). We can now derive the response equation which relates these quantities.

\subsection{The response matrix}

Using the bi-orthogonality condition, we can invert eq.~\eqref{eq:Projection_perturbing_potentials} to get
\begin{equation}
    a_p(t) = - \!\!\int\!\! \rd \bx \, \rho^{\rs}(\bx, t) \, \psi^{(p)*}(\bx).
    \label{eq:Projection_coefficient}
\end{equation}
Since the \DF\@ perturbation is related to the density response in the system through
\begin{equation}
    \rho^{\rs}(\bx, t) = \!\!\int\!\! \rd \bv \, f(\bx, \bv, t),
\end{equation}
equation~\eqref{eq:Projection_coefficient} can be developed as  equation~\eqref{eq:ap}.

Now, the \DF\@ perturbation, $f$, can itself be developed in angular Fourier elements as in eq.~\eqref{eq:Fourier_transform_angles}, and the integration variables can be canonically changed from $\rd \bx \rd \bv$ to $\rd \bJ \rd \bT$, with a Jacobian equal to 1 owing to phase space volume conservation. We then have
\begin{align}
\nonumber
    a_p(t) = & - \sum_{\bn} \!\!\int\!\! \rd \bJ \, f_{\bn}(\bJ, t) \bigg(\!\!\int\!\! \rd \bT \, \re^{ - \ri \bn \cdot \bT}  \, \psi^{(p)}(\bJ, \bT) \bigg)^* \\
= & - (2 \pi)^3 \sum_{\bn} \!\!\int\!\! \rd \bJ \, f_{\bn}(\bJ, t) \, \psi_{\bn}^{(p)*}(\bJ),
\end{align}
where the last expression was obtained thanks to eq.~\eqref{eq:Fourier_transform_angles}. 

While for now, we only used the definition of the projection coefficient and some field equations, let us include the dynamics through the \CBE, and in particular eq.~\eqref{eq:f_Fourier_time_integral}, to get
\begin{align}
    \nonumber
    a_p(t) \!\! = \!\! - \ri (2 \pi)^3 \!\! \int_0^t \!\! \rd \tau \sum_{\bn} & \!\!\int\!\! \rd \bJ \, \re^{- \ri \, \bn \cdot \bO \, (t - \tau)} \, \bn \!\cdot\! \frac{\partial F}{\partial \bJ} \\ 
    & \times \psi_{\bn}^{(p)*\!}(\bJ) \, \Delta H_{\bn} (\bJ, \tau).
    \label{eq:Coeff_ap_psi_DeltaH}
\end{align}

To proceed further, we develop the perturbation to the Hamiltonian in more detail. Starting from eq.~\eqref{eq:Perturbing_Hamiltonian}, let us first expand all potentials ($\psi^{\re}$ and $\psi^{\rs}$) and densities ($\rho_1$ in eq.~\ref{eq:Centre_acceleration}) in the potential-density basis elements. This gives
\begin{equation}
    \Delta H(\bx, t) \!=\! \sum_q (a_q + b_q) \bigg[ \psi^{(q)}(\bx) + \bx \cdot \!\! \int \!\! \frac{G \, \rd \bx}{|\bx|^2} \, \rho^{(q)}\!(\bx) \,\mathbf{e_r} \bigg],
    \nonumber
\end{equation}
where the time dependence is fully borne by the projection coefficients $a_q$ and $b_q$. Defining the new set of functions of eq.~\eqref{eq:Definition_inertial_term}, the angular Fourier transform of the perturbing Hamiltonian is given by
\begin{equation}
    \Delta H_{\bn}(\bJ, t) = \sum_q (a_q + b_q) \big[ \psi^{(q)}_{\bn}(\bJ) + \phi^{(q)}_{\bn}(\bJ) \big].
    \label{eq:Perturbing_Hamiltonian_Fourier}
\end{equation}

Replacing in eq.~\eqref{eq:Coeff_ap_psi_DeltaH} with eq.~\eqref{eq:Perturbing_Hamiltonian_Fourier}, we get
\begin{equation}
    a_p(t)  = \!\! \int_0^t \!\! \rd \tau \, \sum_q \bM_{pq}(t - \tau) \big[a_q(\tau) + b_q(\tau)\big],
    \label{eq:Response_time_integral_sum}
\end{equation}
where the response matrix is defined as in eq.~\eqref{eq:Response_Matrix}. Equation~\eqref{eq:Response_time_integral_sum} is also another version of eq.~\eqref{eq:Response_time_integral}. 

\section{The Clutton-Brock \\ Bi-orthogonal basis}
\label{app:BOB}

In this appendix, we detail our choice of basis functions, which was first constructed by \cite{CluttonBrock1973}. The potential and density elements of this basis are given by eq.~\eqref{eq:Basis_general_form} with
\begin{subequations}
\label{eq:Clutton_basis}
\begin{equation}
    U_n^{\ell}(r) = A_n^{\ell} \frac{(r/\Rb)^{\ell}}{(1 + (r/\Rb)^2)^{\ell + 1/2}} C_n^{(\ell + 1)} (\chi)
\end{equation}
\begin{equation}
     D_n^{\ell}(r) = B_n^{\ell} \frac{(r/\Rb)^{\ell}}{(1 + (r/\Rb)^2)^{\ell + 5/2}} C_n^{(\ell + 1)} (\chi),
\end{equation}
\end{subequations}
where $\Rb$ is the basis scale radius, $C_n^{(\alpha)}$ are the Gegenbauer polynomials, and the renormalised radius is given by
\begin{equation}
    \chi = \frac{(r/\Rb)^{2} - 1}{(r/\Rb)^{2} + 1}.
\end{equation}
In eq.~\eqref{eq:Clutton_basis}, we defined the normalisation constants 
\begin{subequations}
\begin{equation}
    A_n^{\ell} = - \sqrt{\frac{G}{K^{\ell}_n \, \Rb}} \, 2^{2 \ell + 3} \, \ell! \, \sqrt{\frac{(n + \ell + 1) \, n!}{(n + 2\ell + 1)!}},
\end{equation}
\begin{equation}
    B_n^{\ell} = - \frac{K_n^{\ell}}{4 \pi G \, \Rb^2} \, A_n^{\ell},
\end{equation}
\end{subequations}
where $K_n^{\ell}$ is defined as
\begin{equation}
    K_n^{\ell} = 4 n (n + 2 \ell + 2) + (2 \ell + 1) (2 \ell + 3).
\end{equation}
The radial basis is therefore defined for $0 \leq n \leq \nmax$.

\section{The Baes-van Hese equilibrium distribution function}
\label{app:Baes}

For the choice of anisotropic phase space \DFs\@ with Hernquist density, we relied on the work of \cite{Baes2007} (their eqs.~(92) and~(93)). More specifically, we focused on the particular case of spheres with a constant anisotropy parameter $\beta$, so that the \DF\@ is given by
\begin{align}
\nonumber
    F(E,L) = & \frac{1}{(2 \pi)^{5/2}} \frac{1}{(G M a)^{3/2}} \bigg(-\frac{a E}{G M} \bigg)^{5/2 - 2 \beta} \\
    \nonumber
    \times & \frac{\Gamma (5 - 2 \beta)}{\Gamma(1 - \beta) \Gamma(\tfrac{7}{2} - \beta)} \bigg( -\frac{L^2}{2 a^2 E} \bigg)^{-\beta} \\
    \times & _2 F_1 \bigg( 5 - 2 \beta, 1-2\beta, \frac{7}{2} - \beta, -\frac{a E}{G M} \bigg),
    \label{eq:Baes_DF}
\end{align}
where $a$ stands for the scale radius of the \MW, $\aMW$, and $M$ for the total mass of the \MW, $\MMW$. In order to produce a non-negative \DF, the anisotropy parameter is restricted to $\beta \leq 0.5$ \citep[see][]{An+2006}. Note that, in the case where the \DF\@ only represents the stellar halo (Sections~\ref{sec:Response_Anisotropic_Halo} and~\ref{sec:Discussion}), this \DF\@ should merely be rescaled by the factor $\Mtot / \MMW$, without rescaling the energy nor the angular momentum (but see Section~\ref{subsec:Stellar_Halo_Response}).

An extra step is required to consider this \DF\@ as a function of the actions, so that it can be input in eq.~\eqref{eq:Response_Matrix}. Indeed, once the spherically symmetric potential is specified, a bijective relation exists between the sets of conserved quantities that are $(E,L)$ and $(J_r, L)$. In practice, all quantities that enter eq.~\eqref{eq:Response_Matrix} are actually computed in a third set of conserved quantities, the peri- and apocentres $(\rp, \ra)$, which make the coordinate transforms more straightforward. Some technical details of these coordinate transforms are given in Appendix~\ref{app:Matrix_computation}.

\section{Computation of the response matrix}
\label{app:Matrix_computation}

For a spherical, non-rotating mean-field \DF, the formula giving the matrix method can be simplified. We perform these simplifications in the following section. Later on, we describe in some detail the numerical techniques we developed for the matrix computation, and validate the implementation by recovering unstable modes from the literature.

\subsection{Matrix of a spherical, non-rotating equilibrium}
\label{subsec:Spherical_matrix}

Here, we consider the special case were $\Omega_3 = 0$ and $\partial F / \partial L_z = 0$, i.e.\@ that of a spherical, non-rotating system. Let us first use a derivation from \cite{TremaineWeinberg1984} for the Fourier transformed basis functions,
\begin{equation}
    \psi^{(p)}_{\bn} (\bJ) \! = \! \delta_{m^{p}}^{n_{3}} \ri^{m^{p} - n_{2}} Y_{\ell^{p}}^{n_{2}}\! (\tfrac{\pi}{2} , 0) R^{\ell^{p}}_{n_{2} m^{p}}\! (\beta) W^{\tbn}_{\ell^{p} n^{p}}\! (\tbJ),
    \label{eq:Fourier_transformed_basis}
\end{equation}
where $\bn = (n_1, n_2, n_3)$ is the resonance vector associated to each Fourier coefficient, $\beta$ is the inclination angle of the orbit associated to $\bJ$, defined so that $\cos(\beta) = L_z/L$, and $\tbJ = (J_r, L)$, $\tbn = (n_1, n_2)$. Additionally, the rotation matrix ${R_{n m}^{\ell} (\beta) }$ is defined as
\begin{align}
\nonumber
R_{n m}^{\ell} (\beta) \! & = \! \sum_{t} (-1)^{t} \frac{\sqrt{(\ell \! + \! n)! \, (\ell \! - \! n)! \, (\ell \! + \! m)! \, (\ell \! - \! m)!}}{(\ell \! - \! m \! - \! t)! \, (\ell \! + \! n \! - \! t)! \, t! \, (t \! + \! m \! - \! n)!}  \\
& \times \big[ \! \cos (\beta / 2) \! \big]^{2 \ell + n - m - 2 t}  \big[\! \sin (\beta / 2)\! \big]^{2 t + m - n} ,
\label{eq:Rotation_matrix}
\end{align}
where the sum over $t$ is restricted to the values such that the arguments
of the factorials are positive, i.e.\ ${ t_{\rm min} \!\leq\! t \!\leq\! t_{\rm max} }$,
with ${ t_{\rm min} \!=\! \text{Max} [0 , n - m] }$ and
${ t_{\rm max} \!=\! \text{Min} [\ell - m , \ell + n] }$.
In eq.~\eqref{eq:Fourier_transformed_basis}, the Fourier-transformed
``in-plane'' radial coefficients ${ W^{\tbn}_{\ell n} (\tbJ) }$ are defined as
\begin{equation}
W^{\tbn}_{\ell n} (\tbJ) = \frac{1}{\pi} \!\! \int \!\! \rd \theta_{1} \, U_{n}^{\ell} (r (\theta_{1})) \, \cos [n_{1} \theta_{1} + n_{2} (\theta_{2} - \xi)] ,
\label{eq:Definition_W_basis}
\end{equation}
which are real for real radial basis functions.
In this integral, the radial dependence of the angles $\theta_{1}$ and ${ (\theta_{2} - \xi ) }$ is given by 
\begin{subequations}
\label{eq:Angles_radial_integral}
\begin{equation}
    \theta_{1} = \Omega_{1} \!\! \int_{\mC} \!\! \rd r \, \frac{1}{\sqrt{2 (E - \psi_0 (r)) - L^{2} / r^{2}}} ,
\end{equation}
\begin{equation}
    \theta_{2} - \xi = \!\! \int_{\mC} \!\! \rd r \, \frac{\Omega_{2} - L / r^{2}}{\sqrt{2 (E - \psi_0 (r)) - L^{2} / r^{2}}} ,
\end{equation}
\end{subequations}
where $\xi$ is the angle between the ascending node and the current position, measured in the orbit plane along the orbital motion, $E, L$ are the energy and angular momentum of the orbit, and $\mC$ is the integration contour going from the pericentre $\rp$ up to the current position ${ r = r (\theta_{1}) }$ along the radial oscillation.

In order to simplify eq.~\eqref{eq:Response_Matrix}, we also need to decompose the inertial term $\phi_{\bn}^{(p)}(\bJ)$. Using eq.~\eqref{eq:Definition_inertial_term}, we have
\begin{equation}
    \phi_{\bn}^{(p)}(\bJ) \! = \! G \bigg(\!\! \int \!\!\! \frac{\rd \bT}{(2\pi)^3} \, \bx \, \re^{- \ri \, \bn \cdot \bT} \! \bigg) \!\! \cdot \!\! \bigg(\!\! \int \!\! \frac{\rd \bx}{|\bx|^2} \, \rho^{(p)}(\bx) \, \mathbf{e_r} \! \bigg).
    \label{eq:Fourier_transformed_inertial_term}
\end{equation}
Let us first focus on the left-hand integral, which we will rewrite $\bx_{\bn}(\bJ)$, as it is the angular Fourier transform of the position vector. We can rewrite $\bx$ as a Cartesian vector in terms of spherical coordinates $(r, \theta, \phi)$ as
\begin{equation}
    \bx = r \, \mathbf{e_r} = r \, \sqrt{\frac{2 \pi}{3}} 
    \begin{pmatrix}
    Y_1^{-1}(\theta, \phi) - Y_1^{1}(\theta, \phi) \\
    \ri \, \big[Y_1^{-1}(\theta, \phi) + Y_1^{1}(\theta, \phi)\big] \\
    \sqrt{2} \, Y_1^{0}(\theta, \phi)
    \end{pmatrix}.
\end{equation}
In each of the Cartesian directions, we therefore have to perform an angular Fourier transform of a function which is separable in terms of a linear combination of spherical harmonics $Y$, times a function which depends on the radius only (in the present case, $r$ itself). This is precisely the context where eq.~\eqref{eq:Fourier_transformed_basis} can be applied. As a result, we have
\begin{align}
    \nonumber 
    \bx_{\bn}(\bJ) = & \sqrt{\frac{2 \pi}{3}} \, Y_1^{n_2}(\tfrac{\pi}{2},0) \, X^{\tbn}(\tbJ) \, \ri^{-n_2} \\
    & \times \begin{pmatrix}
    - \ri \big[ \delta_{-1}^{n_3} \, R_{n_2 -1}^1(\beta) + \delta_{1}^{n_3} \, R_{n_2 1}^1(\beta) \big] \\
    \delta_{-1}^{n_3} \, R_{n_2 -1}^1(\beta) - \delta_{1}^{n_3} \, R_{n_2 1}^1(\beta) \\
    \sqrt{2} \, \delta_0^{n_3} \, R_{n_2 0}^{1}(\beta)
    \end{pmatrix},
    \label{eq:Fourier_transformed_position}
\end{align}
where we defined new in-plane radial coefficients as
\begin{equation}
    X^{\tbn} (\tbJ) = \frac{1}{\pi} \!\! \int \!\! \rd \theta_{1} \, r (\theta_{1}) \, \cos [n_{1} \theta_{1} + n_{2} (\theta_{2} - \xi)].
    \label{eq:Definition_X_integral}
\end{equation}

Similarly, we can express the right-hand integral of eq.~\eqref{eq:Fourier_transformed_inertial_term} as a Cartesian vector in terms of spherical harmonics. Given the harmonic dependence of the basis functions (see eq.~\ref{eq:Basis_general_form}), as well as the orthogonality of the spherical harmonics, most of the basis functions will yield a vanishing integral. The only non-zero terms give
\begin{equation}
    \!\! \int \!\! \frac{\rd \bx}{|\bx|^2} \rho^{(p)}\!(\bx) \, \mathbf{e_r} = \sqrt{\frac{2\pi}{3}} \, \delta_{\ell^p}^1 \! 
    \begin{pmatrix}
    \! \delta_{m^p}^{-1} - \delta_{m^p}^1 \!\! \\
    \! - \ri \, (\delta_{m^p}^{-1} + \delta_{m^p}^1) \!\! \\
    \! \sqrt{2} \, \delta_{m^p}^0 \!\!
    \end{pmatrix}
    \! d_{n^p},
    \label{eq:Integral_basis_inertial_term}
\end{equation}
where we defined the radial integral of the basis functions over the whole radial range as
\begin{equation}
    d_n = \! \int_0^{\infty} \!\! \rd r \, D_{n}^{1}(r).
\end{equation}

If we now perform the scalar product of eqs.~\eqref{eq:Fourier_transformed_position} and~\eqref{eq:Integral_basis_inertial_term}, which are both written in cartesian coordinates, we have
\begin{align}
\nonumber
    \phi_{\bn}^{(p)}(\bJ) \!=\! \frac{4\pi G d_{n^p}}{3} \delta_{\ell^p}^1 \delta_{m^p}^{n_3} & \ri^{m^p - n_2} Y_{1}^{n_2}(\tfrac{\pi}{2},0) \\
    & \times R_{n_2 m^p}^1(\beta) X^{\tbn}(\tbJ).
    \label{eq:Fourier_inertial_term_final}
\end{align}
The fact that this term accounts for the translation of the reference frame is recovered, since it is only present in dipolar harmonics $\ell^p = 1$.

We can now use eqs.~\eqref{eq:Fourier_transformed_basis} and~\eqref{eq:Fourier_inertial_term_final} to simplify eq.~\eqref{eq:Response_Matrix}. In the latter, the dependence on the third action, $L_z$, is only borne by the rotation matrices $R(\beta)$. We can therefore make use of their orthogonality relation, 
\begin{equation}
\int_{0}^{\pi} \!\!\!\! \rd \beta \, \sin (\beta) \, R^{\ell^{p}}_{n_{2} m^{p}} (\beta) \, R^{\ell^{q}}_{n_{2} m^{p}} (\beta) = \delta_{\ell^{p}}^{\ell^{q}} \, \frac{2}{2 \ell^{p} + 1} .
\label{eq:Rotation_matrices_orthogonality}
\end{equation}
Once this simplification is performed, we end up with the final form of the response matrix for spherical, non-rotating systems,
\begin{equation}
    \bM_{pq}(t) = \delta_{m^p}^{m^q} \, \delta_{\ell^p}^{\ell^q} \, \sum_{\tbn} C_{\ell^p}^{n_2} \, P_{\ell^p n^p n^q}^{\tbn}(t),
    \label{eq:Response_matrix_spherical}
\end{equation}
where we define the coefficients $C_{\ell}^{n}$ as
\begin{equation}
    C_{\ell}^{n} = - 2 \, \ri \, (2 \pi)^3 \, \frac{\big| Y_{\ell}^{n}(\tfrac{\pi}{2},0)\big|^2}{2 \ell + 1},
\end{equation}
and the functions $P_{\ell^p n^p n^q}^{\tbn}(t)$ as
\begin{align}
    \nonumber 
    P_{\ell^p n^p n^q}^{\tbn}(t) = \!\! \int \! & \rd \tbJ \, L \,\, \tbn\! \cdot \!\frac{\partial F}{\partial \tbJ} \,\, \re^{- \ri \, \tbn \cdot \tbO \, t} \, W_{\ell^p n^p}^{\tbn}(\tbJ)  \\
    \times \, & \big[ W_{\ell^p n^q}^{\tbn}(\tbJ) \!+\! \delta_{\ell^p}^1 \frac{4 \pi G}{3} d_{n^q} X^{\tbn}(\tbJ) \big].
    \label{eq:Definition_function_P}
\end{align}
One can notice the similarity of this equation with eq.~(23) of \cite{Murali1999}, the main difference being our definition of $d_{n^q}$ (their $p_j^{lm}$), which stems from considering the reference frame of the cusp instead of that of the \COM. Interestingly enough, the response matrix element $\bM_{pq}$ is proportional to $\delta_{m^p}^{m^q} \, \delta_{\ell^p}^{\ell^q}$. This means that there is no coupling between different angular harmonics in the system's response: each angular harmonic effect in the response is only induced by the corresponding harmonic cause in the perturber, mediated by the corresponding harmonic term in the matrix. As shown by \cite{Rozier2019}, this characteristic is specific to non-rotating spheres. Let us now detail the numerical methods which we used to compute the response matrix.

\subsection{Numerical methods}

To compute the action space integral of eq.~\eqref{eq:Definition_function_P}, we carefully analysed the different terms of the integrand. A critical feature appears when this integrand is rewritten as $g(\tbJ) \, \re^{\ri \, h(\tbJ)}$. In this form, the integrand is written as a slowly varying function of the actions, $g$, times a fast trigonometric oscillation. The argument of this oscillating term itself, $h$, is also a slowly varying function of the actions. We therefore choose to divide the action space in small surfaces, on which both functions $g$ and $h$ are well approximated by their first order expansion. 

In order to reach a better sampling of action space, we relied on the same change of variables as in \cite{Rozier2019}: the integration variables are changed to $(u,v)$, which are written as functions of the orbits' peri- and apocentres. This change of variables allows for a logarithmic sampling of the orbits that are either close to the host's centre or close to circular, while the other orbits are sampled linearly in terms of peri- and apocentre. This usually leads to a better sampling of the regions where the integrand of eq.~\eqref{eq:Definition_function_P} reaches a significant amplitude. The function $g$ can be redefined to include the transformation's Jacobian, and the resulting functions $g(u,v)$ and $h(u,v)$ are still slowly varying functions of their arguments, as compared to the fast trigonometric oscillation. 

In the end, we compute eq.~\eqref{eq:Definition_function_P} as the sum over a grid on the $(u,v)$ surface of the integral $\aleph(g,\tfrac{\partial g}{\partial u},\tfrac{\partial g}{\partial v},h,\tfrac{\partial h}{\partial u},\tfrac{\partial h}{\partial v})$ defined by
\begin{equation}
    \aleph(a,\! b,\! c,\! d,\! e,\! f) \!=\! \!\! \int\!\!\!\!\!\int_{\! - \tfrac{\Delta u}{2}}^{\! \tfrac{\Delta u}{2}} \!\!\!\! \rd u \rd v \, (a \!+\! b u \!+\! c v) \, \re^{\ri (d + e u + f v)}\!,
\end{equation}
where $g, h$ and their derivatives are evaluated at the centre $(u_0, v_0)$ of each square of side $\Delta u$. This integral can be renormalised as
\begin{equation}
    \aleph(a,\! b,\! c,\! d,\! e,\! f) \!=\! \Delta u^2 \, a \, \re^{\ri d} \, \aleph_{\mathrm D}(\!\tfrac{b \Delta u}{a}\!,\! \tfrac{c \Delta u}{a}\!,\! \tfrac{e \Delta u}{d}\!,\! \tfrac{f \Delta u}{d}\!),
\end{equation}
where the normalised integral is defined as
\begin{equation}
    \aleph_{\mathrm D}(b,c,e,f) = \!\!\int\!\!\!\!\!\int_{-\tfrac{1}{2}}^{\tfrac{1}{2}} \!\!\! \rd x \, \rd y \, (1 + b x + c y) \, \re^{\ri \, (e x + f y)}.
\end{equation}
Finally, we found an analytical expression for this last integral as
\begin{align}
    \nonumber 
    \aleph_{\mathrm D}(b,c,e,f) = & \sinc (\tfrac{e}{2}) \, \sinc (\tfrac{f}{2}) \\
    \nonumber
    & - \ri \, \frac{b}{e} \, \sinc (\tfrac{f}{2}) (\cos(\tfrac{e}{2}) - \sinc(\tfrac{e}{2})) \\
    & - \ri \, \frac{c}{f} \, \sinc (\tfrac{e}{2}) (\cos(\tfrac{f}{2}) - \sinc(\tfrac{f}{2})),
\end{align}
where $\sinc(x) = \sin(x)/x$.

Let us now explain in more detail how we evaluate the functions $g$, $h$ and their partial derivatives. In general, most of the functions involved in $g$ and $h$ can be expressed as functions of $(\rp, \ra)$, the orbit's peri- and apocentre. Such functions can later be considered as functions of $(u,v)$, owing to the analytical relations $\rp(u)$ and $\ra(u,v)$ \citep[see][]{Rozier2019}. In particular, the energy (required in the phase space \DF\@ $F(E,L)$) and angular momentum are given by
\begin{subequations}
\label{eq:EL_rpra}
\begin{equation}
    E = \frac{\ra^2 \, \psi_0(\ra) - \rp^2 \, \psi_0(\rp)}{\ra^2-\rp^2},
\end{equation}
\begin{equation}
    L = \sqrt{\frac{2(\psi_0(\ra)-\psi_0(\rp))}{\rp^{-2}-\ra^{-2}}},
\end{equation}
\end{subequations}
and the orbital frequencies are given by
\begin{subequations}
\label{eq:Frequencies_rpra}
\begin{equation}
    \Omega_1 = \bigg[\frac{1}{\pi} \! \int_{\rp}^{\ra} \!\! \rd r \, \frac{1}{\sqrt{2 (E - \psi_0(r)) - L^2/r^2}} \bigg]^{-1},
\end{equation}
\begin{equation}
    \Omega_2 = \frac{\Omega_1}{\pi} \! \int_{\rp}^{\ra} \!\! \rd r \, \frac{L/r^2}{\sqrt{2 (E - \psi_0(r)) - L^2/r^2}}.
\end{equation}
\end{subequations}
Owing to these relations, the functions $g$ and $h$ can be computed, as well as their partial derivatives through explicit analytical expressions. In the case of both $g$ and its partial derivatives, a special treatment should be mentioned in the computation of $W$, $X$ and their partial derivatives (see eqs.~\ref{eq:Definition_W_basis} and~\ref{eq:Definition_X_integral}). Indeed, these functions a priori involve nested integrals of the form
\begin{equation}
    \int \!\! \rd r \, S\big[r, \theta_1(r), \big(\theta_2 - \xi\big)(r) \big],
\end{equation}
where $\theta_1(r)$ and $\big(\theta_2 - \xi\big)(r)$ themselves are integrals (see eq.~\ref{eq:Angles_radial_integral}). Besides, the integrands involved are unbound at the edges of the integration region, which could be the source of issues when performing derivatives. To cure these two problems, we first regularise the integrals at their edges by applying the same effective anomalies as in \cite{Henon1971} \citep[see also][]{Rozier2019}. Then, following \cite{Rozier2019}, instead of directly computing the nested integrals, we transform the problem into the single integration of a multi-component vector. These tricks allow us to compute $W$, $X$, as well as their partial derivatives, as simple well-posed integrals using an RK4 integration scheme.

\subsection{Validation of the implementation}
\label{subsec:ROI_Validation}

\begin{figure}
\centering
    \includegraphics[width=0.45\textwidth]{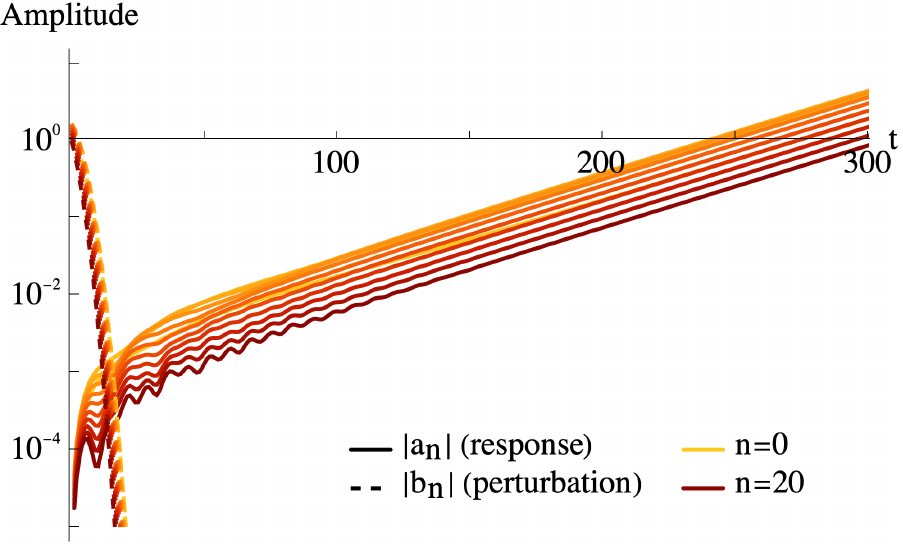} 
    \caption{Time evolution of the norm of the coefficients of the perturber $|b_n|$ (dashed lines) and the response $|a_n|$ (full lines). Different colors indicate different values of $n$. Note that the different $|b_n|$ values were artificially shifted in order to avoid overlapping. Despite the early disappearance of the external perturber, the system spontaneously develops an exponentially growing mode at the rate $\eta = 0.0245$.}
\label{fig:ROI_Coeffs}
\end{figure}

In order to validate our implementation of the response matrix, we present here the response of a radially anisotropic isochrone sphere. In \cite{Saha1991}, it is shown that such a sphere undergoes a radial orbit instability. We therefore compute the response matrix for a spherical system with  an isochrone potential \citep{BinneyTremaine2008}, and an Osipkov-Merritt \DF\@ \citep{Osipkov1979,Merritt1985} with a transition radius from the isotropic centre to the radially anisotropic outskirts taken at $\Ra = 1.0 \, b$, where $b$ is the scale radius of the isochrone potential. Since the instability is expected to emerge as a quadrupolar ($\ell = 2$) mode, we focus on this component of the response matrix and we perturb the system with a potential presenting the corresponding level of symmetry. The computation uses 100 basis functions as defined in Appendix~\ref{app:BOB}, with a scale radius $\Rb = 20 \, b$, and a maximum radial resonance number of $n_{1 \max} = 5$. In this section, all quantities are rescaled so that $G = \Mtot = b = 1$, where $\Mtot$ is the total mass of the sphere. 

By definition, the unstable mode is a property of the background sphere, as a consequence it is independent of the perturber we apply. We therefore arbitrarily choose an axisymmetric perturber ($m=0$), and instead of choosing a potential for the perturber which would later be projected onto the functional basis to get the vector $\bb$, we directly fix the value of the vector as well as its time evolution. For simplicity, we choose to give equal values to all vector coefficients, with a rapid cut-off in time of the form $\re^{- t^2 / 40}$. The time evolution of these coefficients is shown in Fig.~\ref{fig:ROI_Coeffs}. 

Figure~\ref{fig:ROI_Coeffs} also shows the time evolution of the coefficients of the response vector $\ba$, when the response matrix formalism is applied according to eq.~\eqref{eq:Linear_response_final} and evolved up to $t = 300$. Although the perturber is steeply cut-off after $t \sim 10$, it has clearly excited an instability which later grows at an exponential rate: indeed, above $t \sim 25$, all coefficients of $\ba$ grow at the same exponential pace. When we compute the common slope of these curves, we can estimate the growth rate of the identified instability to be $\eta = 0.0245$, which favourably compares to the computation from \cite{Saha1991} at $\eta = 0.025$. In addition to the norm of these coefficients, we measured their phase angle, and did not detect any variation of it. This indicates that the mode is not oscillating, which is also the conclusion of \cite{Saha1991}.

\begin{figure}
\centering
    \includegraphics[width=0.45\textwidth]{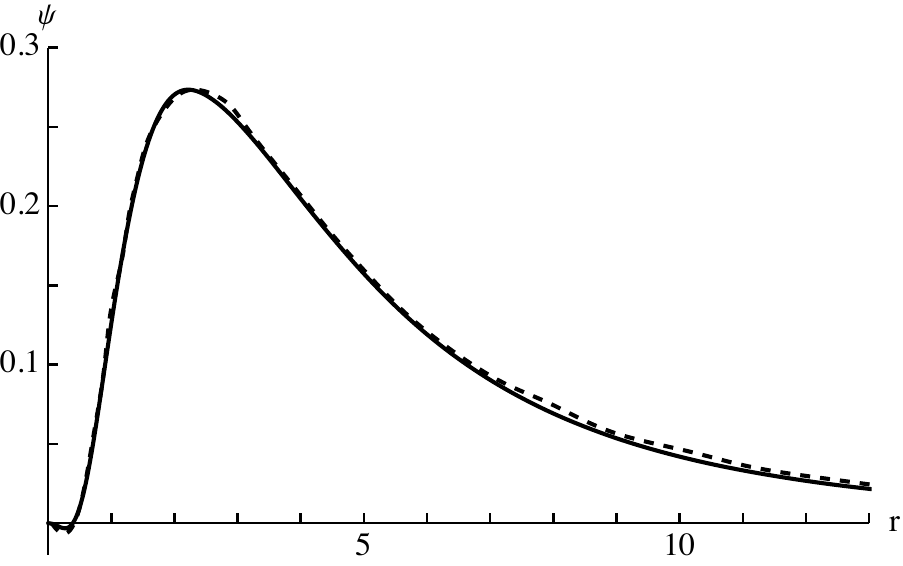} 
    \caption{Radial profile of the instability's potential, as computed by \cite{Saha1991} (dashed line) and by our method (full line). The shapes are very similar, which further validates our implementation.}
\label{fig:ROI_Potential}
\end{figure}

Focusing on the spatial shape of the instability, Fig.~\ref{fig:ROI_Potential} compares the radial profile of the mode's potential from our computation at $t=300$ to the same measurement from \cite{Saha1991} (both using an arbitrary normalisation of the potential's amplitude). The two profiles look very much alike, despite the use of different computation parameters as well as different detection methods (time space in our case vs. frequency space in theirs).

All in all, this comparison suggests that our algorithm is valid for our purposes.

\section{Optimised operations with the response matrix}
\label{app:Matrix_operations}

Equation~\eqref{eq:Linear_response_final} involves the inversion of a large matrix, as well as its product with a large vector. In order to perform these operations, we developed fast algorithms which take into account their specificities. 

As defined by eq.~\eqref{eq:Full_time_matrix}, the matrix $\ubM$ is a block-triangular-Toeplitz matrix. Such a matrix is defined by the following particular shape,
\begin{equation}
    \begin{pmatrix}
    \bM_0 & & & & \\
    \bM_1 & & & \scalebox{1.5}{$\mathbf{0}$} & \\
    \bM_2 & & \ddots & & \\
    \vdots & \ddots & & & \\
    \bM_K & \cdots & \bM_2 & \bM_1 & \bM_0
    \end{pmatrix},
    \label{Toeplitz_matrix}
\end{equation}
where the diagonal blocks are noted $\bM_i$. One can easily show that the inverse of such a matrix is also a block-triangular-Toeplitz matrix. This inverse can be computed by the following recurrence. Let us first coin $\bA_i$ the blocks of the inverse matrix. The first block is straightforwardly given by $\bA_0 = \bM_0^{-1}$, where this inversion is computed using a standard matrix inversion scheme. Then, for $1 \leq i \leq K$, the matrix $\bA_i$ is computed thanks to the relation
\begin{equation}
    \bA_i = - \bM_0^{-1} \sum_{k=0}^{i-1} \bM_{k + 1} \bA_{i - 1 - k}.
\end{equation}
One can easily show that such an algorithm indeed yields the inverse of the original matrix. Note that, when inverting $\ubI - \ubM$, the inversion of the first term is straightforward, since it is equal to the identity itself.

A similar algorithm can be developed for the product of such a block-triangular-Toeplitz matrix with a vector. When the matrix~\eqref{Toeplitz_matrix} is multiplied with a vector defined by the stacked sub-vectors $\bb_0, \cdots, \bb_K$, then the stacked sub-vectors of the product, which we note $\bc_0, \cdots, \bc_K$ can be computed via
\begin{equation}
    \bc_i = \sum_{k=0}^{i} \bM_k \bb_{i-k}.
\end{equation}

\end{document}